\documentclass[preprint,preprintnumbers,amsmath,amssymb,floatfix,nofootinbib,superscriptaddress]{revtex4}
\usepackage[dvips]{graphicx}
\usepackage{url}
\usepackage{subfigure}
\usepackage{color}

\def\be{\begin{equation}}
\def\ee{\end{equation}}
\def\ba{\begin{eqnarray}}
\def\ea{\end{eqnarray}}

\def\mur{\mu_\mathrm{R}}
\def\muf{\mu_\mathrm{F}}
\def\msbar{\overline{\mathrm{MS}}}
\def\POWHEGBOX{{\tt POWHEG~BOX}}
\def\POWHEG{{\tt POWHEG}}
\def\SHERPA{{\tt SHERPA}}

\def\MCNLO{{\tt MC@NLO}}
\def\MadaMCNLO{{\tt MadGraph5\_aMC@NLO}}
\def\PYTHIA{{\tt PYTHIA}}
\def\PYTHIAsix{{\tt PYTHIA6}}
\def\NLOPYT{{\tt POWHEG\!+\!PYTHIA}}

\def\HERWIG{{\tt HERWIG}}
\def\GoSam{{\tt GoSam}}
\def\OpenLoops{{\tt OpenLoops}}
\def\tth{Ht\bar t}
\def\mr{\mathrm}

\def\bbh{Hb\bar b}
\def\bjet{b\!-\!\mr{jet}}

\def\hbb{H+2b}
\def\hb{H+1b}
\def\msbar{\overline{\mr{MS}}}

\begin{document}          
%
%
\title{Higgs boson production in association with $b$~jets in the
  \POWHEGBOX{}}
\author{B.~J\"ager}
\email{barbara.jaeger@itp.uni-tuebingen.de}
\affiliation{Institute for Theoretical Physics, T\"ubingen University,
 Auf der Morgenstelle 14, 72076 T\"ubingen, Germany}
\author{L.~Reina}
\email{reina@hep.fsu.edu}
\affiliation{Physics Department, Florida State University,
Tallahassee, FL 32306-4350, U.S.A.}
\author{D.~Wackeroth}
\email{dow@ubpheno.physics.buffalo.edu}
\affiliation{Department of Physics, SUNY at Buffalo, Buffalo, NY 14260-1500, U.S.A.}

\date{\today}

\begin{abstract}
  The hadronic production of a Higgs boson ($H$) in association with
  $b$~jets will play an important role in investigating the
  Higgs-boson couplings to Standard Model particles during Run II of
  the CERN Large Hadron Collider, and could in particular reveal the
  presence of anomalies in the assumed hierarchy of Yukawa couplings
  to the third-generation quarks.  A very high degree of accuracy in
  the theoretical description of this process is crucial to implement
  the rich physics program that could lead to either direct or
  indirect evidence of new physics from Higgs-boson
  measurements. Aiming for accuracy in the theoretical modeling of
  $H+b$-jet production, we have interfaced the analytic
  Next-to-Leading-Order QCD calculation of $Hb\bar{b}$ production with
  parton-shower Monte Carlo event generators in the \POWHEGBOX{}
  framework.  In this paper we describe the most relevant aspects of
  the implementation and present results for the production of $H+1b$
  jet, $H+2b$ jets, and $H$ with no tagged $b$ jets, in the form of
  kinematic distributions of the Higgs boson, of the $b$~jets, and of
  the non-$b$ jets, at the 13~TeV Large Hadron Collider.  The
  corresponding code is part of the public release of the
  \POWHEGBOX{}.
\end{abstract}
%
\maketitle

\section{Introduction}
\label{sec:introduction}

The production of a Higgs boson with $b$~jets at the CERN Large Hadron
Collider (LHC) can provide essential information on the Higgs-boson
couplings to third generation quarks, in particular to the bottom
quark.  Indeed, all the leading parton-level production processes
($gg\rightarrow Hb\bar{b}$, $q\bar{q}\rightarrow Hb\bar{b}$, and
$bg\rightarrow Hb$) involve a Higgs boson radiated from an external
bottom quark, while at the loop-level the Higgs boson can also
originate from internal loops of both bottom (leading) and top quarks
(see, e.g., the discussion in~\cite{Dawson:2005vi,Wiesemann:2014ioa}).
In the Standard Model, the Higgs-boson couplings to both fermions and
gauge bosons are just proportional to the particles' masses, causing
the Higgs-boson associated production with bottom quarks to be largely
suppressed with respect to the major production mechanisms, like
gluon-gluon fusion (mediated by a loop of top quarks) or vector-boson
fusion and associated production with vector bosons (where the Higgs
boson couples to $W$ or $Z$ bosons). The production with $b$~jets is
then further suppressed by the identification cuts usually placed on
$b$~jets for tagging purposes. This scenario can however be
drastically different if the hierarchy of Higgs-boson Yukawa couplings
is modified by factors that typically enter in models with extended
Higgs sectors, like Two Higgs Doublet Models. In the quest for
unveiling the origin of the breaking of the electroweak gauge
symmetry, the evidence for (or absence of) $H+b$-jet production at
Run~II of the LHC can therefore provide an essential piece of the
puzzle.

In view of its crucial role for the Higgs-boson physics program of Run
II of the LHC, $H+b$-jets production has received quite some attention
in the context of the LHC Higgs Cross Section Working
Group~\cite{Dittmaier:2011ti,Dittmaier:2012vm,Heinemeyer:2013tqa}, and
both ATLAS and CMS have used the $Hb\bar{b}$ production channel in all
major studies to constrain supersymmetric models and other extensions
of the Standard
Model~\cite{Chatrchyan:2013qga,Khachatryan:2014wca,CMS:2015ooa,Khachatryan:2015tra,Aad:2014vgg}.
On the theoretical side, it is essential to control and improve the
accuracy with which we can estimate rates for $H$ production with one
or two $b$~jets, and, with this regard, quite some progress has been
made in the last few years. In the following we will briefly summarize
the main results obtained in this context, while we refer to the
existing literature for more exhaustive explanations and details.

As for all processes involving multiple scales (such as the masses of
the bottom quark or the Higgs boson, $m_b$ and $M_H$, as well as
scales determined by the kinematics of collisions at TeV
center-of-mass energies), the QCD perturbative prediction of
$H+b$-quark production can be affected by the presence, at all orders,
of large corrections proportional to logarithms of ratios of these
mass scales (e.~g., $\log(m_b/Q)$ where indicatively one can assume
$Q\sim M_H$ or larger). The occurrence of these enhanced logarithmic
corrections depends on the signature studied as well as on the kinematic
regime considered.  It can be shown that these corrections can be
reabsorbed in the perturbative definition of a bottom-quark parton
distribution function, and from this observation originates the
prescription to calculate processes like $H+b$ quarks in a 5 Flavor
Scheme (5FS) where the $b$ quark is treated as a light parton and can
appear in the initial state. At fixed perturbative order this is an
alternative to the usual 4 Flavor Scheme (4FS) calculation where only
four light-flavor parton densities are assumed while the bottom quark
is treated as massive and can only appear in the final state. The
production of $H+1b$~jet can be induced at tree level by
$bg\rightarrow bH$ in the 5FS, and by $q\bar{q},gg\rightarrow
Hb\bar{b}$ in the 4FS; while the production of $H+2b$~jets can only
arise at lowest order via $q\bar{q},gg\rightarrow Hb\bar{b}$ and is
therefore an unambiguous 4FS prediction.  The two approaches only
correspond to a different reordering of the (same) perturbative
expansion, and their predictions tend to agree better the higher the
perturbative order, showing the expected well behaved convergence of
QCD predictions.  Several discussions of this issue can be found in
the literature, where the general 4FS/5FS formalism is thoroughly
analyzed~\cite{Maltoni:2012pa,Cordero:2015sba} as well as specialized
to the particular case of $H+b$-quark
production~\cite{Campbell:2004pu,Dawson:2005vi,Heinemeyer:2013tqa,Wiesemann:2014ioa}.

Next-to-Leading Order (NLO) QCD corrections to $H+b$~jet production
have been calculated both in the 5FS~\cite{Campbell:2002zm} and in the
4FS~\cite{Dittmaier:2003ej,Dawson:2003kb,Dawson:2004sh}. The first set
of corrections is nowadays part of the NLO QCD 5FS prediction of
$H+1b$~jet~\cite{Campbell:2002zm,mcfm,Heinemeyer:2013tqa}, while the
second set of corrections are included in the NLO QCD 4FS prediction
of both $H+1b$~jet~\cite{Dawson:2004sh} and $H+2b$~jets
~\cite{Dittmaier:2003ej,Dawson:2003kb,Heinemeyer:2013tqa}. NLO QCD
fixed-order results for both $Hb$ and $Hb\bar{b}$ hadronic production
can also be obtained via any of the public NLO automated
tools such as \MadaMCNLO~\cite{Alwall:2014hca},
\GoSam~\cite{Cullen:2014yla}, or \OpenLoops~\cite{Cascioli:2011va}.
Of course, the totally inclusive cross section (with no $b$ tagging)
can be calculated in either scheme, and dedicated studies which
includes up to Next-to-Next-to-Leading-Order (NNLO) QCD corrections
have been presented in the
literature~\cite{Dicus:1998hs,Maltoni:2003pn,Harlander:2003ai,Campbell:2004pu,Dawson:2005vi,Heinemeyer:2013tqa,Wiesemann:2014ioa}.

In order to improve the accuracy of theoretical predictions for total
cross sections and distributions, these NLO fixed-order results need
to be consistently interfaced with parton-shower Monte Carlo
generators like \PYTHIA~\cite{Sjostrand:2006za,Sjostrand:2007gs} and
\HERWIG~\cite{Marchesini:1991ch,Corcella:2000bw} using one of the
methods proposed in the literature, namely
\MCNLO~\cite{Frixione:2002ik,Frixione:2003ei} and
\POWHEG~\cite{Nason:2004rx,Frixione:2007vw,Frixione:2007nw}, as
implemented in specific frameworks like, e.~g.,
\MadaMCNLO~\cite{Hirschi:2011pa,Alwall:2014hca}, the
\POWHEGBOX~\cite{Alioli:2010xd}, and
\SHERPA~\cite{Gleisberg:2008ta}. The implementation of Higgs-boson
production with $b$ quarks in \MadaMCNLO{} has been discussed in
Ref.~\cite{Wiesemann:2014ioa}, where total cross sections have been
given for both inclusive and exclusive production and distributions
have been shown in particular for the inclusive case (no $b$-jet
tagging). In this paper we present the implementation of $H+b$-jet
production in the \POWHEGBOX, based on the 4FS NLO QCD calculation of
$Hb\bar{b}$ hadronic production of Ref.~\cite{Dawson:2003kb}.  While
Ref.~\cite{Wiesemann:2014ioa} considers both the 5FS and a 4FS cases,
we will only consider the 4FS case since we aim at presenting in
particular results for both $H+1b$~jet and $H+2b$~jets in the same
framework. We note that the implementation of $b$-initiated processes
in a NLO QCD parton-shower Monte Carlo is still being studied and, to
our knowledge, it is not routinely available in any of the
aforementioned frameworks. Hence our decision to only implement the
4FS case. The details of the implementation will be presented in
Section~\ref{sec:powheg-box}. Results for $H+1b$-jet, $H+2b$-jet,
and $H$ with no tagged $b$ jet will be given in
Section~\ref{sec:results}, using a specific setup, for the purpose of
illustrating the kind of studies that are now possible within the
\POWHEGBOX{} framework. Our conclusion are presented in
Section~\ref{sec:conclusions}.

\section{Implementation}
\label{sec:powheg-box}
The implementation of the $\bbh$ process in the framework of the
\POWHEGBOX{} can be performed along the same lines as the related
$\tth$ process that has been considered in
Ref.~\cite{Hartanto:2015uka}. While the \POWHEGBOX{} package provides
all process-independent building blocks, it requires a list of all
independent flavor structures for the tree-level contributions at
Leading Order (LO) and NLO, the Born and real-emission amplitudes
squared, the finite parts of the virtual contributions, the color- and
spin-correlated amplitudes squared, and a parametrization of the
phase space for the Born process. The flavor structures and tree-level
amplitudes can most conveniently be generated with the help of a
tool based on {\tt
  MadGraph~4}~\cite{Stelzer:1994ta,Alwall:2007st} that is provided in
the \POWHEGBOX{}. The virtual contributions for the $pp\to \bbh$
process are extracted from the NLO-QCD calculation of
\cite{Dawson:2003kb} and adapted to the format required by the
\POWHEGBOX{}. All building blocks are implemented in the 4FS, i.e. no
contributions from incoming bottom quarks are taken into account and
the bottom-quark mass is always considered to be non-zero.

While at LO and in the real-emission contributions only diagrams
including a $\bbh$ coupling emerge, in the virtual corrections also
loop diagrams with a $\tth$ coupling contribute. These are fully taken
into account in the representative results discussed in this work. The
user of the \POWHEGBOX{}~implementation can choose to de-activate the
contributions including a top-quark Yukawa coupling via a switch in
the input file. This allows to rescale separately the two
contributions as necessary to calculate $\bbh$ production in, for
instance, supersymmetric extensions of the Standard Model, as discussed
in detail in~\cite{Dawson:2005vi}, where also a rescaling prescription
is provided.

We note that, contrary to the case of $\tth$ production, where the
heavy-quark mass is typically renormalized in the on-shell
renormalization scheme, in the case of $\bbh$ production the
renormalized bottom-quark mass is often defined in the $\msbar$
renormalization scheme~\cite{Dawson:2003kb,Dawson:2005vi}.  While both
renormalization schemes are perturbatively equivalent at NLO with
differences only due to higher-order contributions, $\bbh$ production
processes have been found to be quite sensitive to the renormalization
scheme via the bottom-mass dependence of the overall bottom-quark
Yukawa coupling. Indeed, as higher-order corrections beyond the
one-loop level are partly taken care of, physical observables are
often found to exhibit a better perturbative behavior when the
$\msbar$ scheme is used for the the bottom-quark Yukawa
coupling. Taking this into account, the results presented below
have been obtained in the $\msbar$ scheme.
When using the \POWHEGBOX{} implementation of the $\bbh$ process,
however, the user is free to choose either the on-shell or the
$\msbar$ renormalization scheme for the bottom-quark mass that enters
the Yukawa coupling by setting the respective parameter in the input
file.

Although in principle not necessary for obtaining finite results,
technical cuts at the generation level can help to improve the
performance of the Monte-Carlo integration. For the computation of
observables with identified $b$~jets we therefore recommend the use of
a small cut on the transverse-momentum of the bottom quarks
(e.g.\ $p_T^\mr{cut}=0.1$~GeV) when the phase-space integration is
performed. We have checked that final results for the respective scenarios in
Section~\ref{sec:results} do not change when generation cuts of
$p_T^\mr{cut}=0.1$~GeV or $p_T^\mr{cut}=1$~GeV are imposed compared to
the case where no generation cuts are applied.

In order to verify the \POWHEGBOX{} implementation of the $\bbh$
process, we have performed a detailed comparison of cross sections and
distributions at LO and NLO as obtained in the \POWHEGBOX{} with the
fixed-order code of \cite{Dawson:2003kb}, and found full agreement for
all considered observables. In addition we have successfully compared our results
to those of Ref.~\cite{Wiesemann:2014ioa}.

\section{Results}
\label{sec:results}
The code we developed is available from the webpage of the \POWHEGBOX{}
project, {\tt http://powhegbox.mib.infn.it/}. With this version of the
code the user is free to study $\bbh$ production at a hadron collider
in a customized setup. Here, we wish to discuss representative results
for $\bbh$ production at the LHC with a center-of-mass energy of
$\sqrt{s}=13$~TeV. We use the four-flavor MSTW2008 set of parton
distribution functions \cite{Martin:2009iq,Martin:2010db} as
implemented in the LHAPDF library~\cite{Whalley:2005nh}, with the
associated value of $\alpha_s$, a Higgs-boson mass of $m_H=125$~GeV,
and a top-quark mass of $m_t = 173$~GeV. The on-shell mass of the
bottom quark is set to $m_b^\mr{OS}=4.75$~GeV, resulting in an
$\msbar$ mass of $\overline{m}_b(\overline{m}_b)=4.34$~GeV at NLO~QCD.

For the renormalization ($\mur$) and factorization ($\muf$) scales we
consider two options: first we use a fixed scale,
\be
\label{eq:fix}
\mu_0 = \frac{m_H+2m_b}{4}\,,
\ee
and second a dynamical scale, 
\be
\label{eq:dyn}
\mu_0 = \frac{1}{4}\sum_i \sqrt{m_i^2 + p_{T,i}^2}\,,
\ee
where the summation runs over the masses and transverse momenta of the
Higgs boson and the partons in the final state of the fixed-order
calculation. To assess the scale dependence of our predictions, we
vary the renormalization and factorization scales, $\mur=\xi \mu_0$
and $\muf=\xi \mu_0$ simultaneously in the range $\xi=1/2$ to $\xi=
2$.

We have matched the fixed-order NLO calculation with {\tt
  PYTHIA-6.4.25}. In order to be able to focus our discussion of the
NLO+\PYTHIA{} results on genuine parton-shower effects, we did not
activate multi-parton interactions, underlying event effects, or
decays of the Higgs boson in the Monte-Carlo program, although each of
these effects could in principle be accounted for by setting the
respective parameters in \PYTHIA{}.

In our numerical analysis we consider the two scenarios with a Higgs boson
produced in association with one or two identified $b$~jets, as well as the case with
no tagged $b$~jets. Jets of any type are reconstructed with the
anti-$k_T$ algorithm as implemented in the {\tt FASTJET}
package~\cite{Cacciari:2011ma}, with $R=0.5$. A jet that contains
either a bottom quark or antiquark, or a $B$ meson, is considered a
$b$~jet. For our $\hbb$-jet analysis, we require at least two $b$~jets
with a minimum transverse momentum in the central region of
pseudorapidity,
\be
p_T(\bjet) > 25~\mr{GeV}\,,\quad
|\,\eta(\bjet)| < 2.5\,,
\ee
while for the $\hb$-jet analysis only one identified $b$~jet
fulfilling the above criteria is required. We do not impose any cuts
on extra jets, unless stated otherwise.

\begin{figure}[tp]
\begin{center}
\begin{tabular}{lr}
\includegraphics[scale=0.7,height=8truecm,width=8truecm,trim=10 0 70 20,clip]{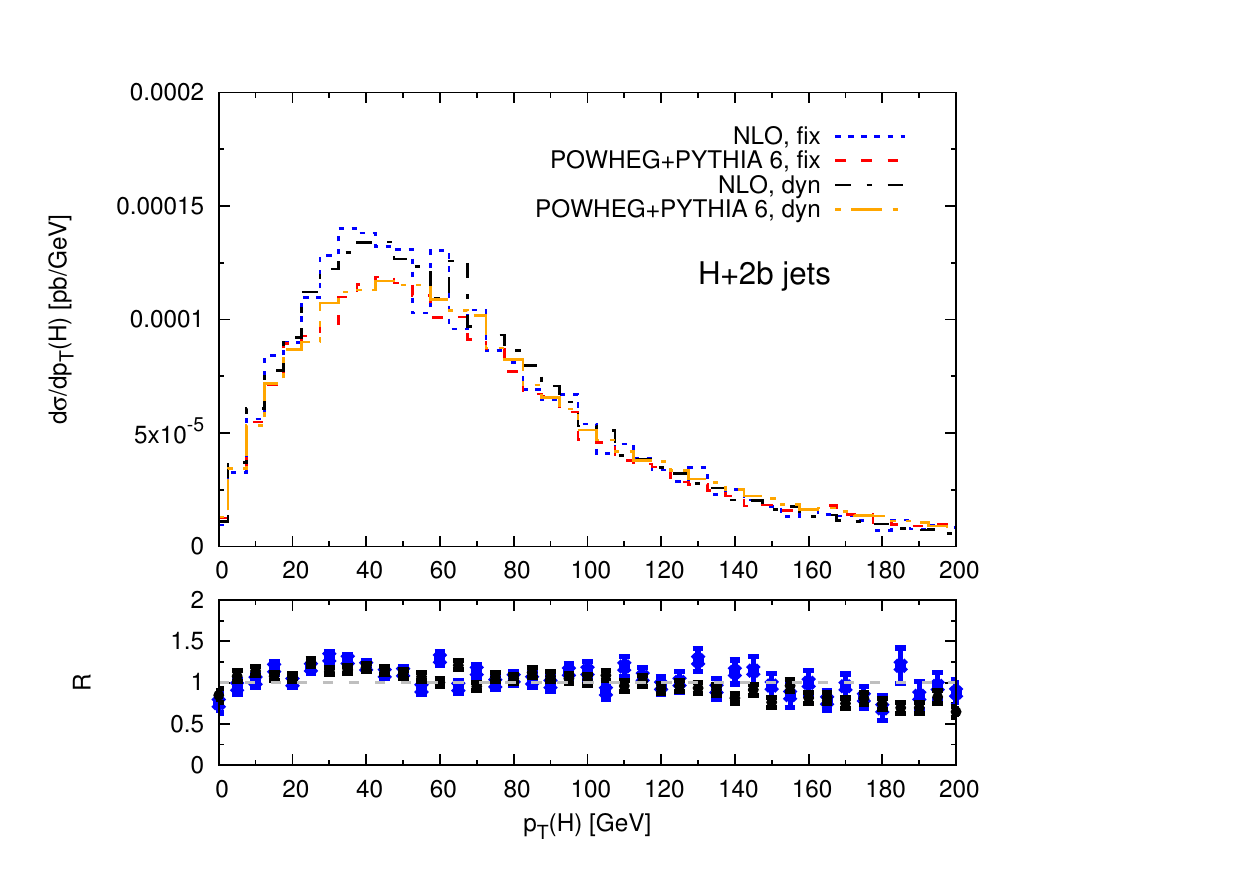} &
\includegraphics[scale=0.7,height=8truecm,width=8truecm,trim=10 0 70 20,clip]{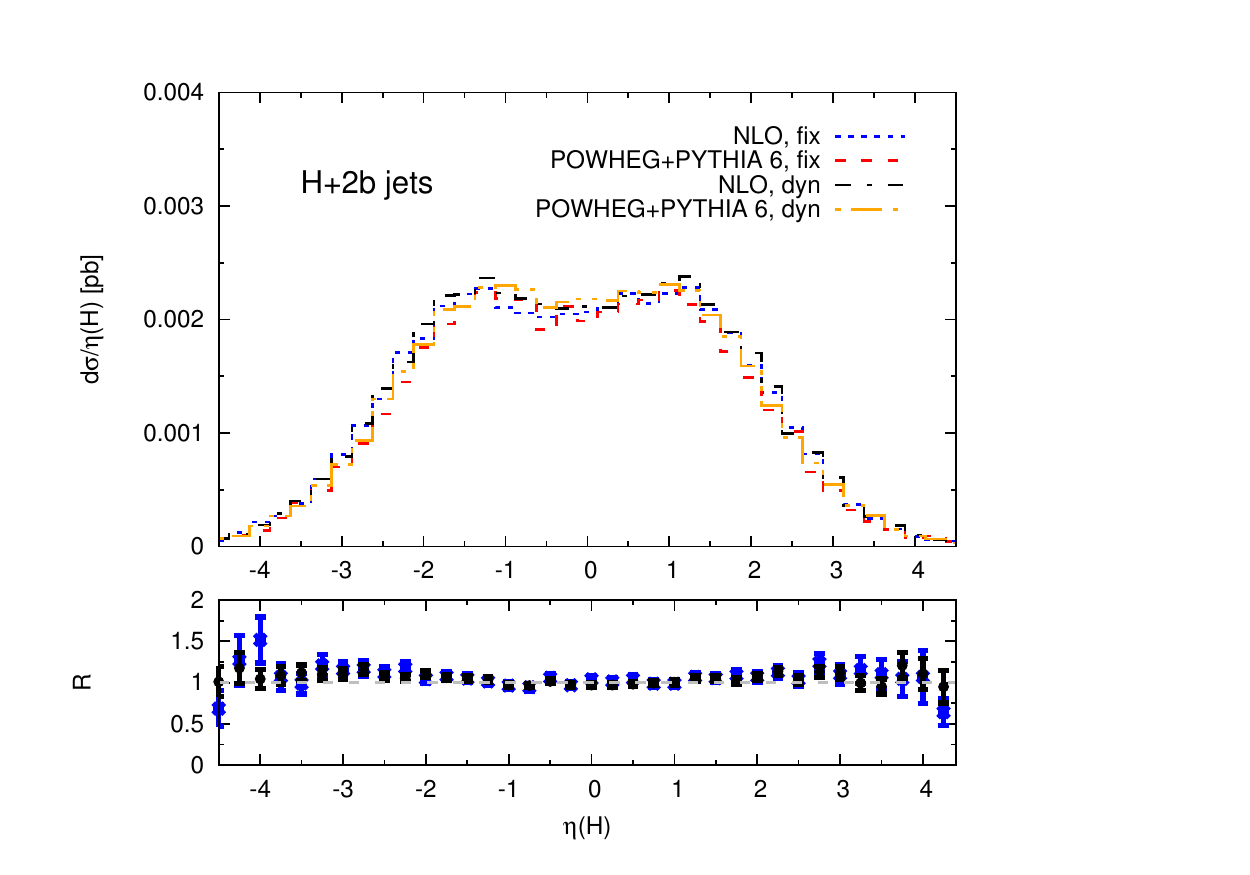}
\end{tabular}
\caption{The $p_T$ (left) and $\eta$ (right) distributions of the
  Higgs boson in $H+2 b$-jet production at NLO-QCD with no parton
  shower (blue short-dashed line for fixed scale, black
  long-dash-dotted line for dynamical scale), and with parton shower as
  obtained through \POWHEG+\PYTHIAsix{} (red medium-dashed line for
  fixed scale, orange long-dash-dotted line for dynamical scale).  The
  lower panels show the ratios:
  $R=d\sigma(\mr{NLO})/d\sigma(\NLOPYT{})$ for a fixed (blue points)
  and a dynamical (black points) scale, respectively.  The error bars
  indicate the statistical uncertainties of the Monte-Carlo
  integration.}
\label{fig:pteta_h_nlo_pythia6_2b}
\end{center}
\end{figure}

\begin{figure}[tp]
\begin{center}
\begin{tabular}{lr}
\includegraphics[scale=0.7,height=8truecm,width=8truecm,trim=10 0 70 20,clip]{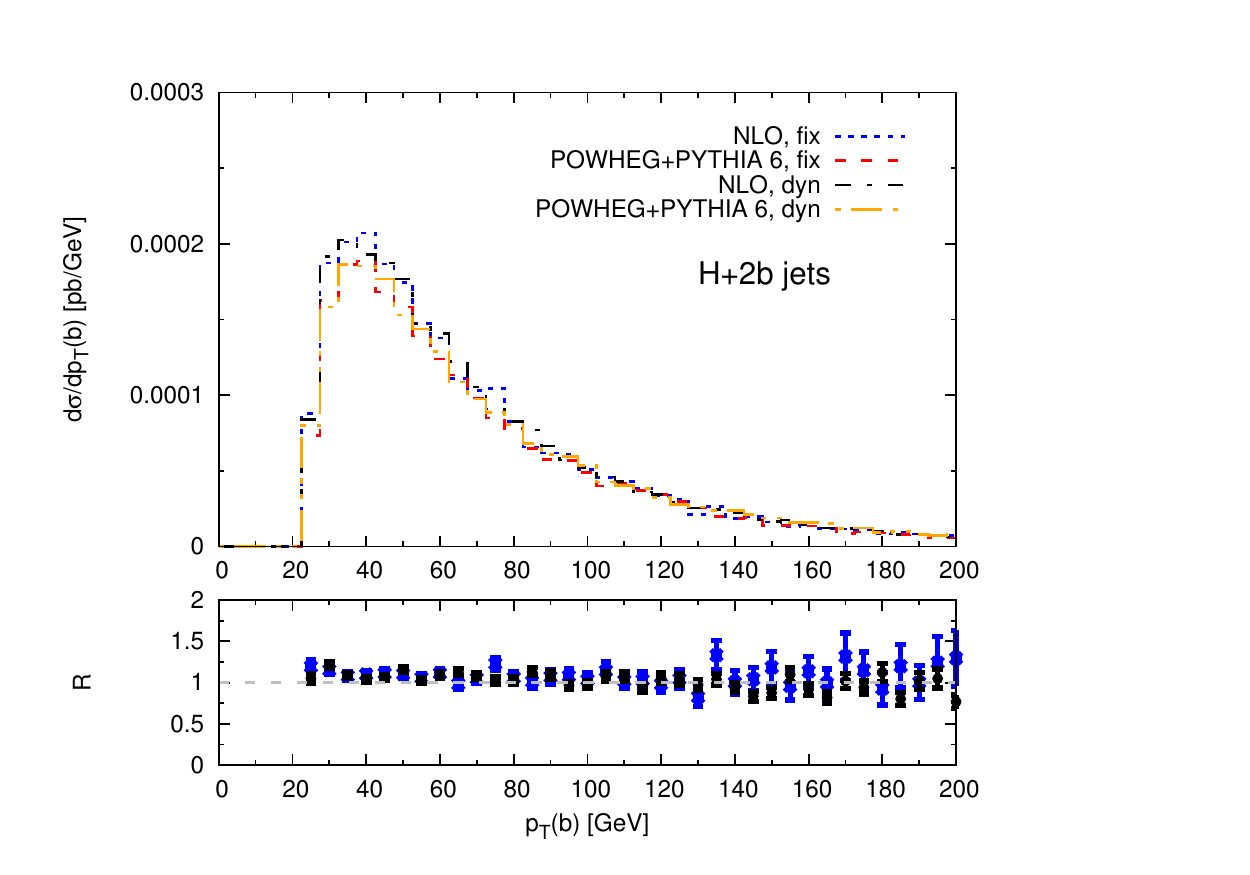} &
\includegraphics[scale=0.7,height=8truecm,width=8truecm,trim=10 0 70 20,clip]{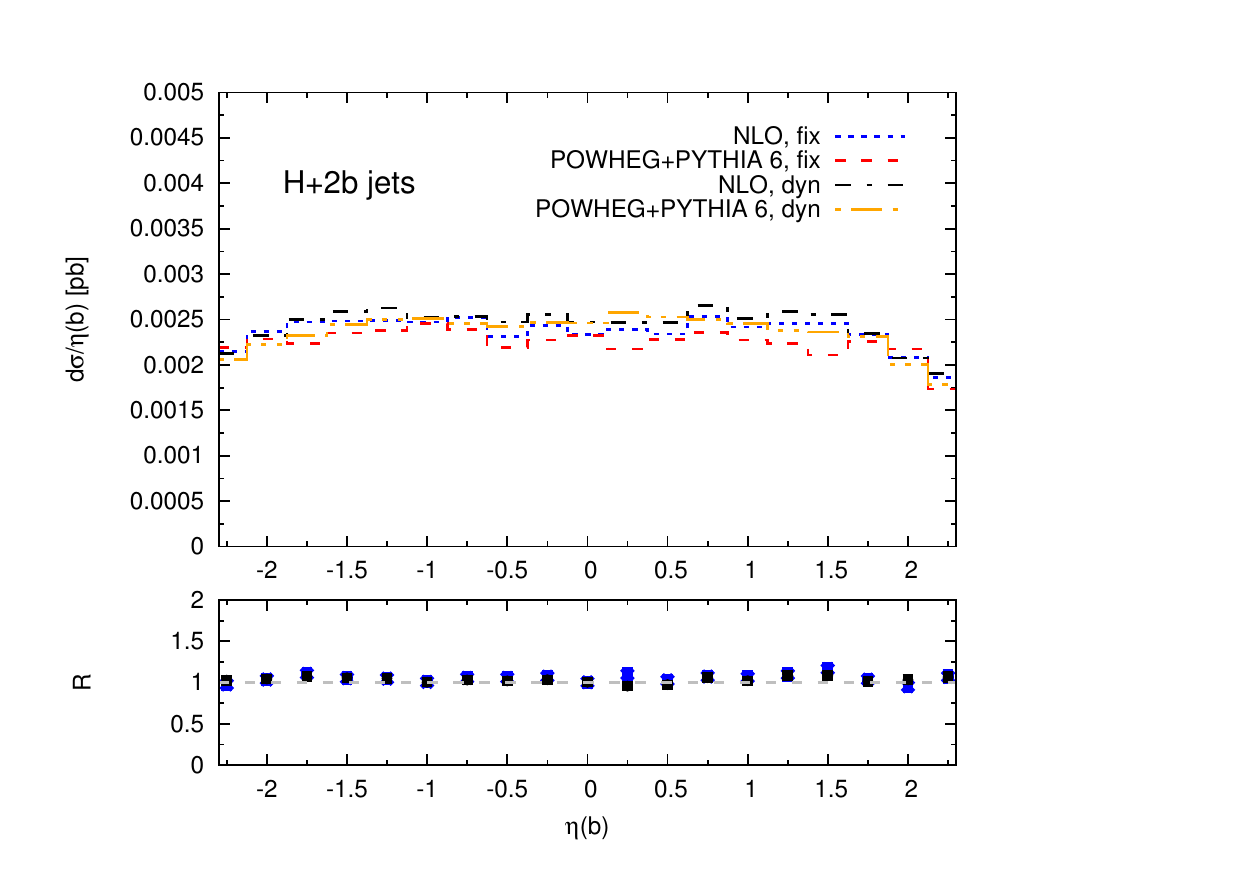}
\end{tabular}
\caption{The $p_T$ (left) and $\eta$ (right) distributions of the
  hardest identified $b$~jet in $H+2 b$-jet production at NLO-QCD with
  no parton shower (blue short-dashed line for fixed scale, black
  long-dash-dotted line for dynamical scale), and with parton shower as
  obtained through \POWHEG+\PYTHIAsix{} (red medium-dashed line for
  fixed scale, orange long-dash-dotted line for dynamical scale).  The
  lower panels show the ratios:
  $R=d\sigma(\mr{NLO})/d\sigma(\NLOPYT{})$ for a fixed (blue points)
  and a dynamical (black points) scale, respectively.  The error bars
  indicate the statistical uncertainties of the Monte-Carlo
  integration.}
\label{fig:pteta_b_nlo_pythia6_2b}
\end{center}
\end{figure}

\begin{figure}[tp]
\begin{center}
\begin{tabular}{lr}
\includegraphics[scale=0.7,height=8truecm,width=8truecm,trim=10 0 70 20,clip]{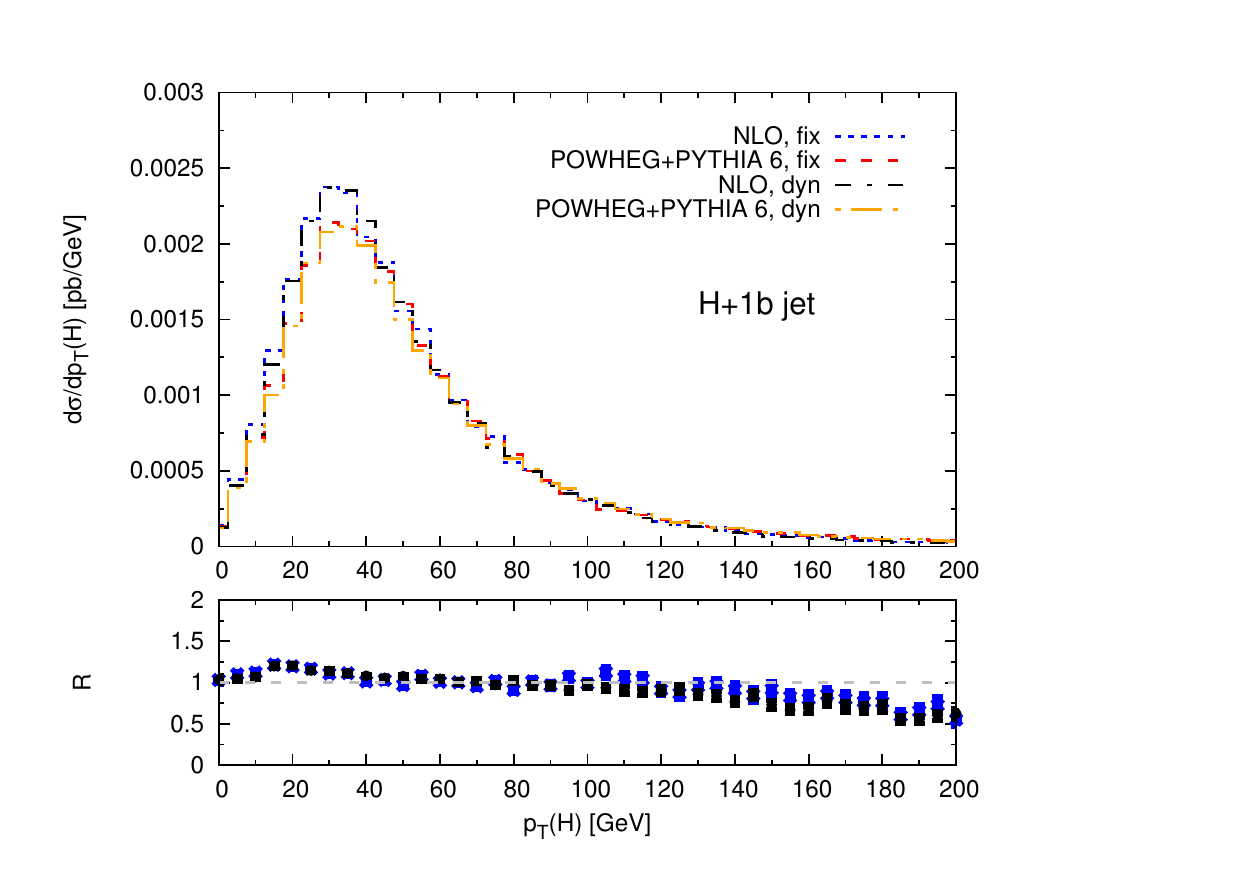} &
\includegraphics[scale=0.7,height=8truecm,width=8truecm,trim=10 0 70 20,clip]{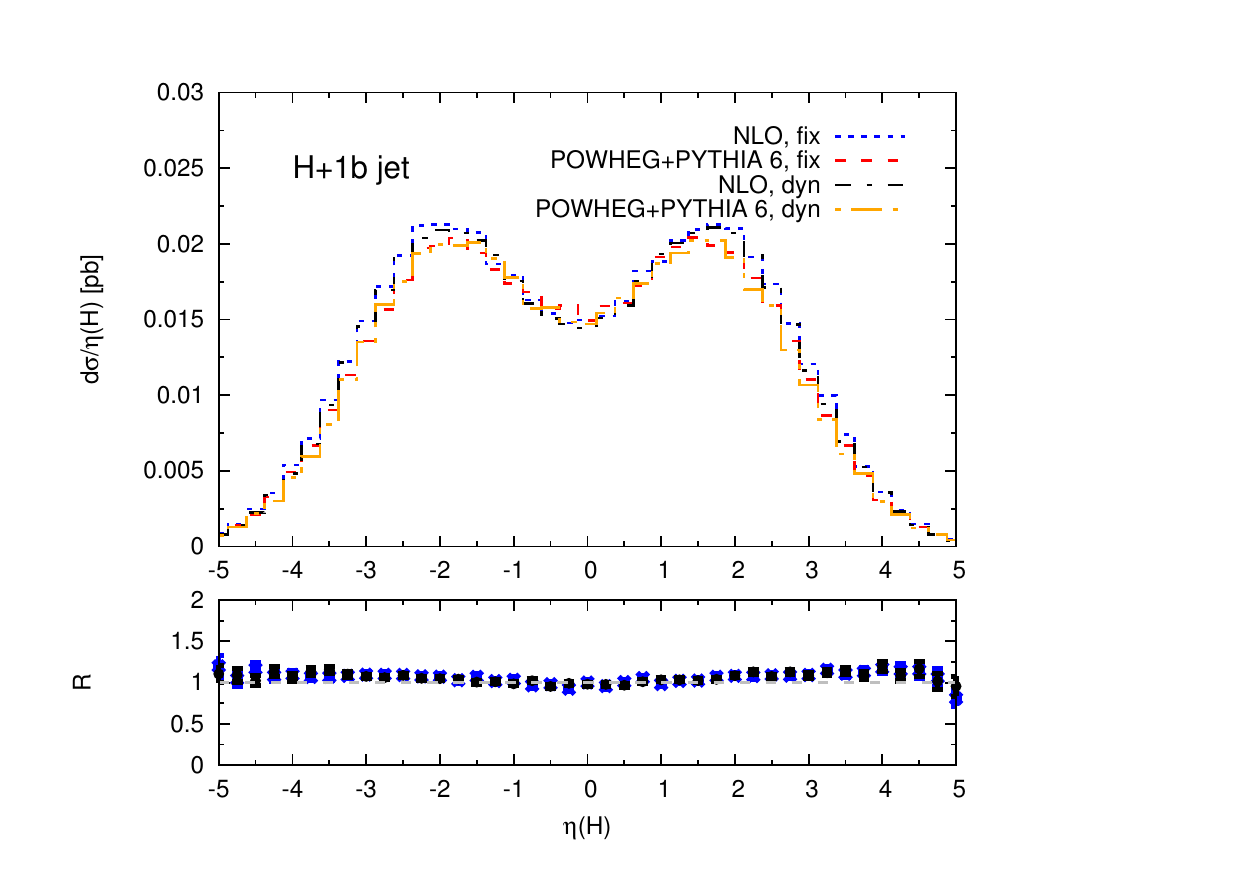}
\end{tabular}
\caption{The $p_T$ (left) and $\eta$ (right) distributions of the
  Higgs boson in $H+1 b$-jet production at NLO-QCD with no parton
  shower (blue short-dashed line for fixed scale, black
  long-dash-dotted line for dynamical scale), and with parton shower as
  obtained through \POWHEG+\PYTHIAsix{} (red medium-dashed line for
  fixed scale, orange long-dash-dotted line for dynamical scale).  The
  lower panels show the ratios:
  $R=d\sigma(\mr{NLO})/d\sigma(\NLOPYT{})$ for a fixed (blue points)
  and a dynamical (black points) scale, respectively.  The error bars
  indicate the statistical uncertainties of the Monte-Carlo
  integration.}
\label{fig:pteta_h_nlo_pythia6_1b}
\end{center}
\end{figure}

\begin{figure}[tp]
\begin{center}
\begin{tabular}{lr}
\includegraphics[scale=0.7,height=8truecm,width=8truecm,trim=10 0 70 20,clip]{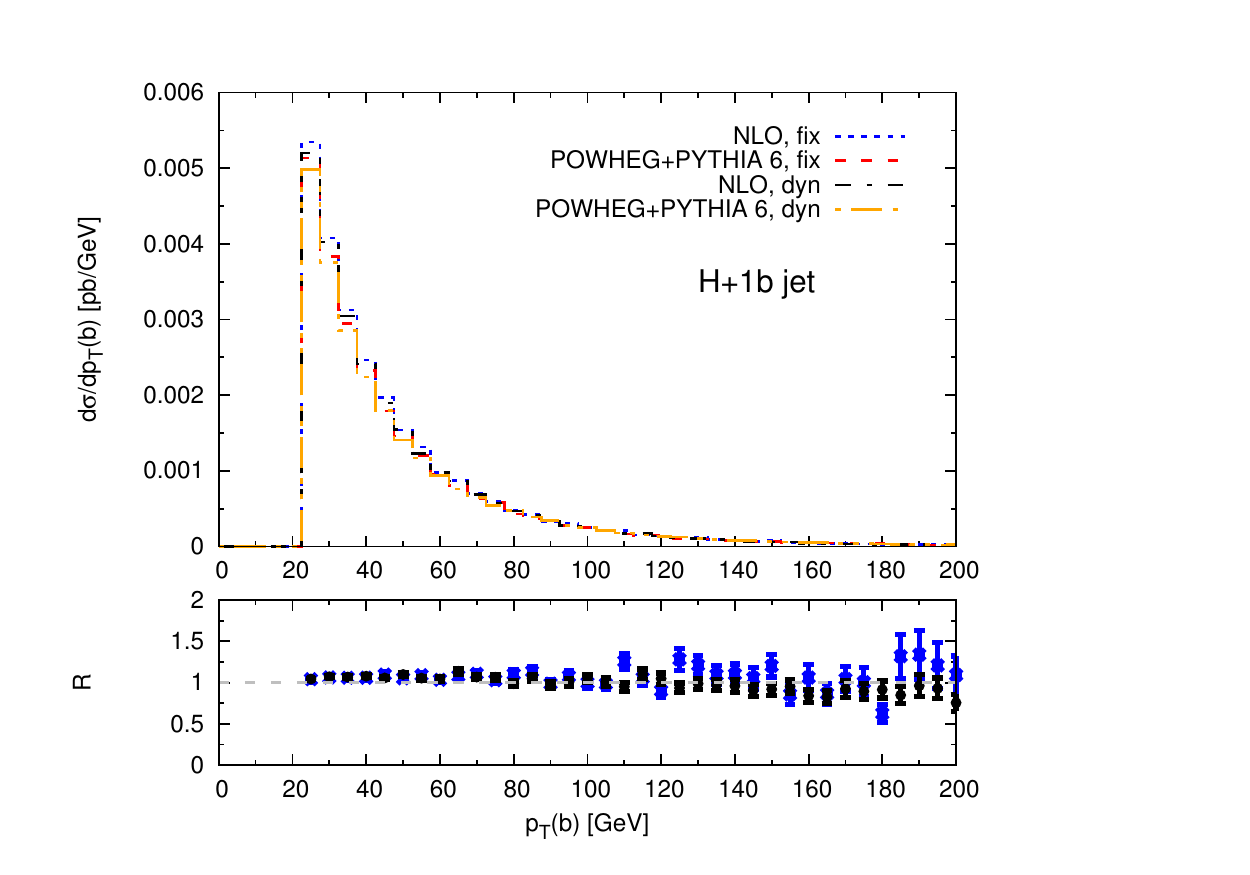} &
\includegraphics[scale=0.7,height=8truecm,width=8truecm,trim=10 0 70 20,clip]{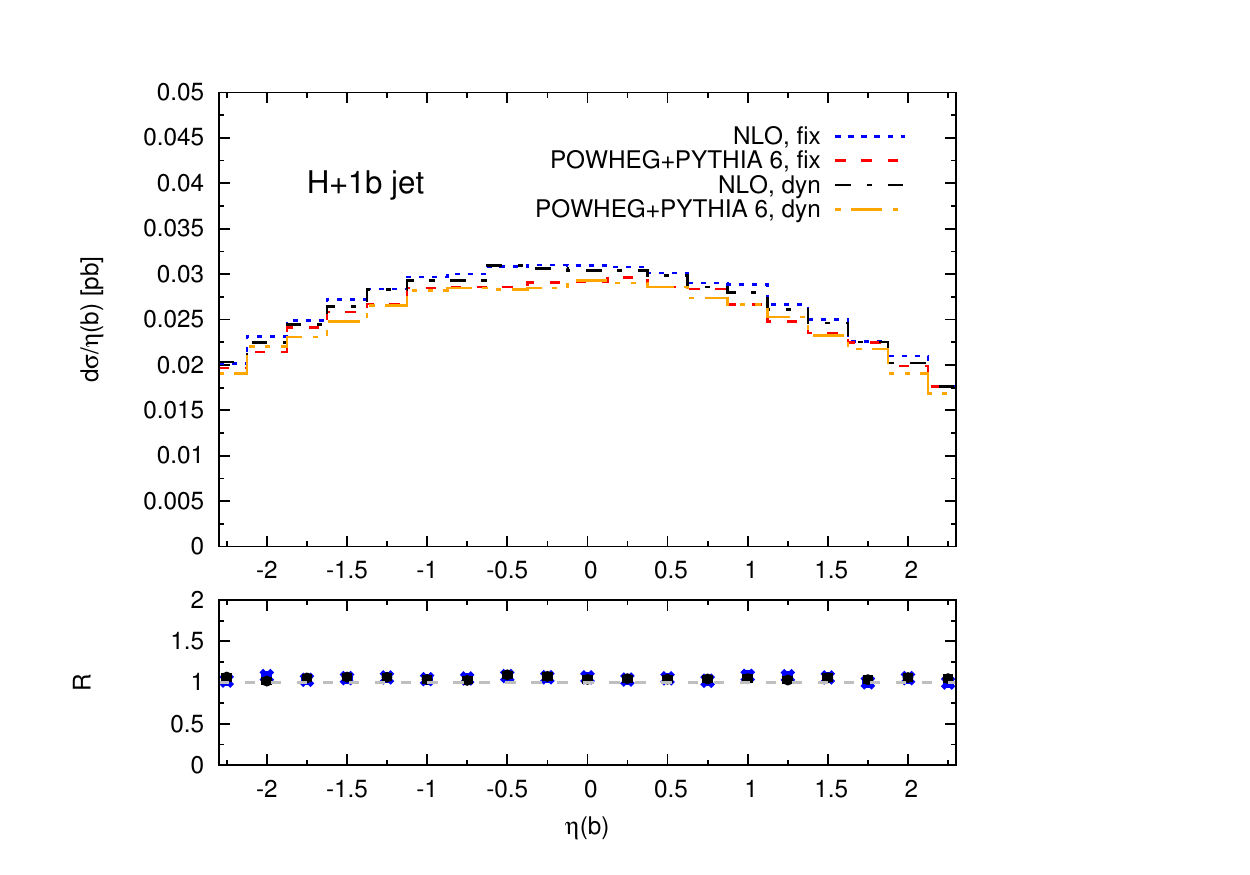}
\end{tabular}
\caption{The $p_T$ (left) and $\eta$ (right) distributions of the
  hardest identified $b$~jet in $H+1 b$-jet production at NLO-QCD with
  no parton shower (blue short-dashed line for fixed scale, black
  long-dash-dotted line for dynamical scale), and with parton shower as
  obtained through \POWHEG+\PYTHIAsix{} (red medium-dashed line for
  fixed scale, orange long-dash-dotted line for dynamical scale).  The
  lower panels show the ratios:
  $R=d\sigma(\mr{NLO})/d\sigma(\NLOPYT{})$ for a fixed (blue points)
  and a dynamical (black points) scale, respectively.  The error bars
  indicate the statistical uncertainties of the Monte-Carlo
  integration.}
\label{fig:pteta_b_nlo_pythia6_1b}
\end{center}
\end{figure}

\begin{figure}[tp]
\begin{center}
\begin{tabular}{lr}
\includegraphics[scale=0.7,height=8truecm,width=8truecm,trim=10 0 70 20,clip]{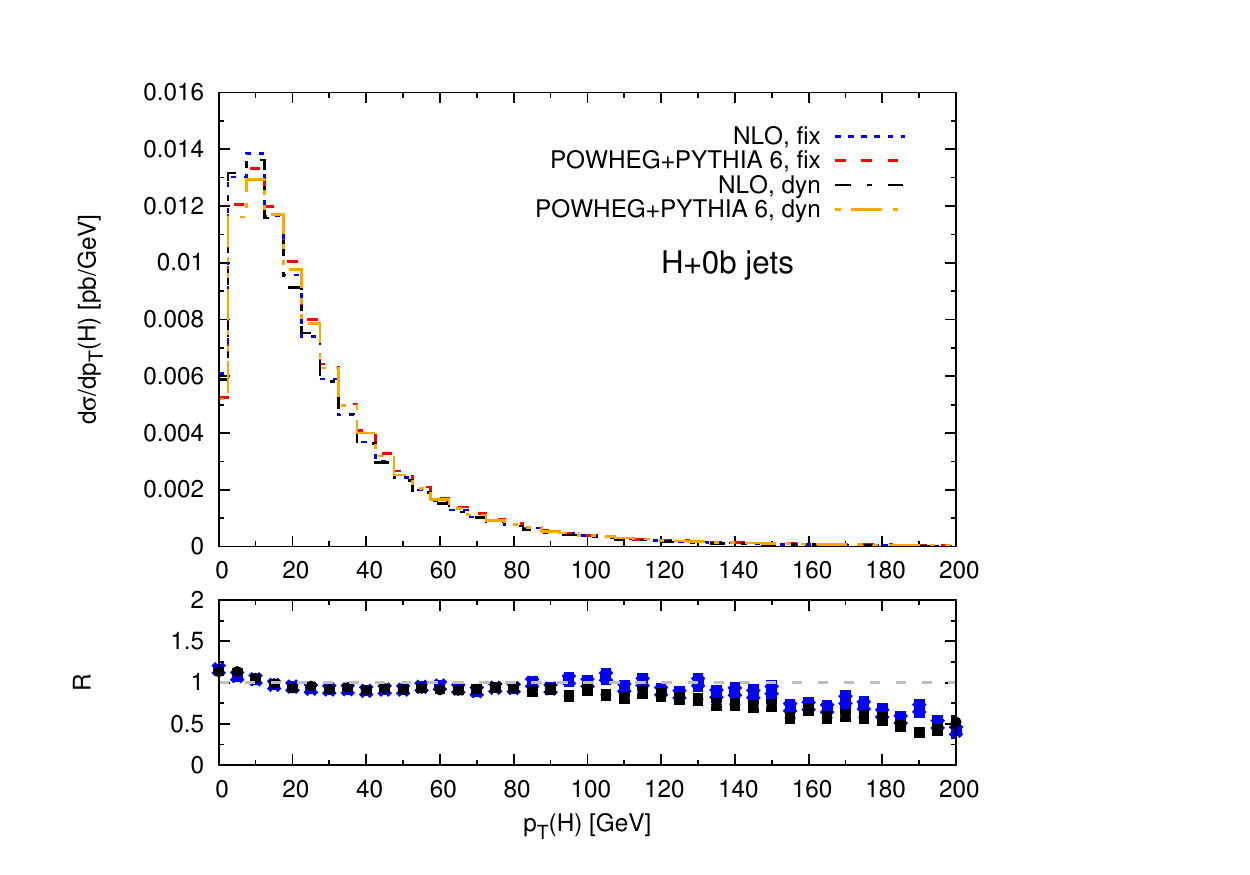} &
\includegraphics[scale=0.7,height=8truecm,width=8truecm,trim=10 0 70 20,clip]{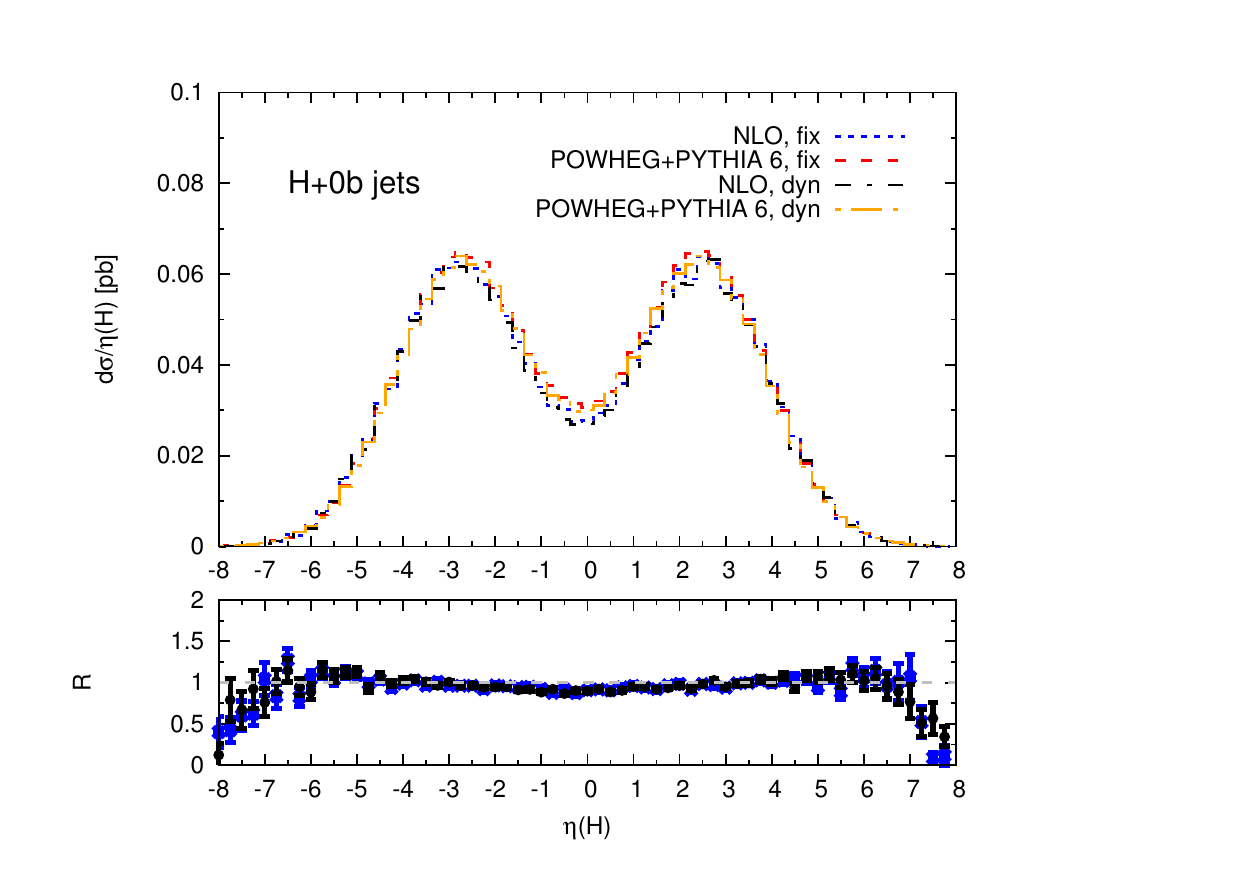}
\end{tabular}
\caption{The $p_T$ (left) and $\eta$ (right) distributions of the
  Higgs boson in the inclusive case at NLO-QCD with
  no parton shower (blue short-dashed line for fixed scale, black
  long-dash-dotted line for dynamical scale), and with parton shower as
  obtained through \POWHEG+\PYTHIAsix{} (red medium-dashed line for
  fixed scale, orange long-dash-dotted line for dynamical scale).  The
  lower panels show the ratios:
  $R=d\sigma(\mr{NLO})/d\sigma(\NLOPYT{})$ for a fixed (blue points)
  and a dynamical (black points) scale, respectively.  The error bars
  indicate the statistical uncertainties of the Monte-Carlo
  integration.}
\label{fig:pteta_h_nlo_pythia6_0b}
\end{center}
\end{figure}

In
Figs.~\ref{fig:pteta_h_nlo_pythia6_2b}-\ref{fig:pteta_b_nlo_pythia6_2b}
we illustrate the impact of the parton shower on the fixed-order NLO
QCD results for the transverse-momentum ($p_T$) and pseudorapidity
($\eta$) distributions of the Higgs boson and the hardest of the
identified $b$~jets in $H+2 b$-jet production, respectively, for both
the fixed- and the dynamical-scale choice $\muf=\mur=\mu_0$ of
Eqs.~(\ref{eq:fix}) and (\ref{eq:dyn}), respectively.  In
Figs.~\ref{fig:pteta_h_nlo_pythia6_1b}-\ref{fig:pteta_b_nlo_pythia6_1b}
we show the corresponding distributions for $H+1 b$-jet production,
and in Fig.~\ref{fig:pteta_h_nlo_pythia6_0b} we show the $p_T(H)$ and
$\eta(H)$ distributions for the inclusive case, i.e. when no $b$~jets
are tagged. We find that parton-shower effects do not significantly
change the fixed-order NLO results of these distributions within the given
statistical uncertainty in most of the kinematic regimes shown. Only
in the Higgs $p_T$ distributions for all signatures, $H+0,1,2 b$~jets,
we find that parton-shower effects decrease (enhance) the NLO results
for small (large) values of $p_T(H)$. These distributions also show
that the effects of the parton shower do not significantly differ for
a fixed- and dynamical-scale choice within their respective statistical
uncertainties.


\begin{figure}[tp]
\begin{center}
\begin{tabular}{lr}
\includegraphics[scale=0.7,height=8truecm,width=8truecm,trim=10 0 70 20,clip]{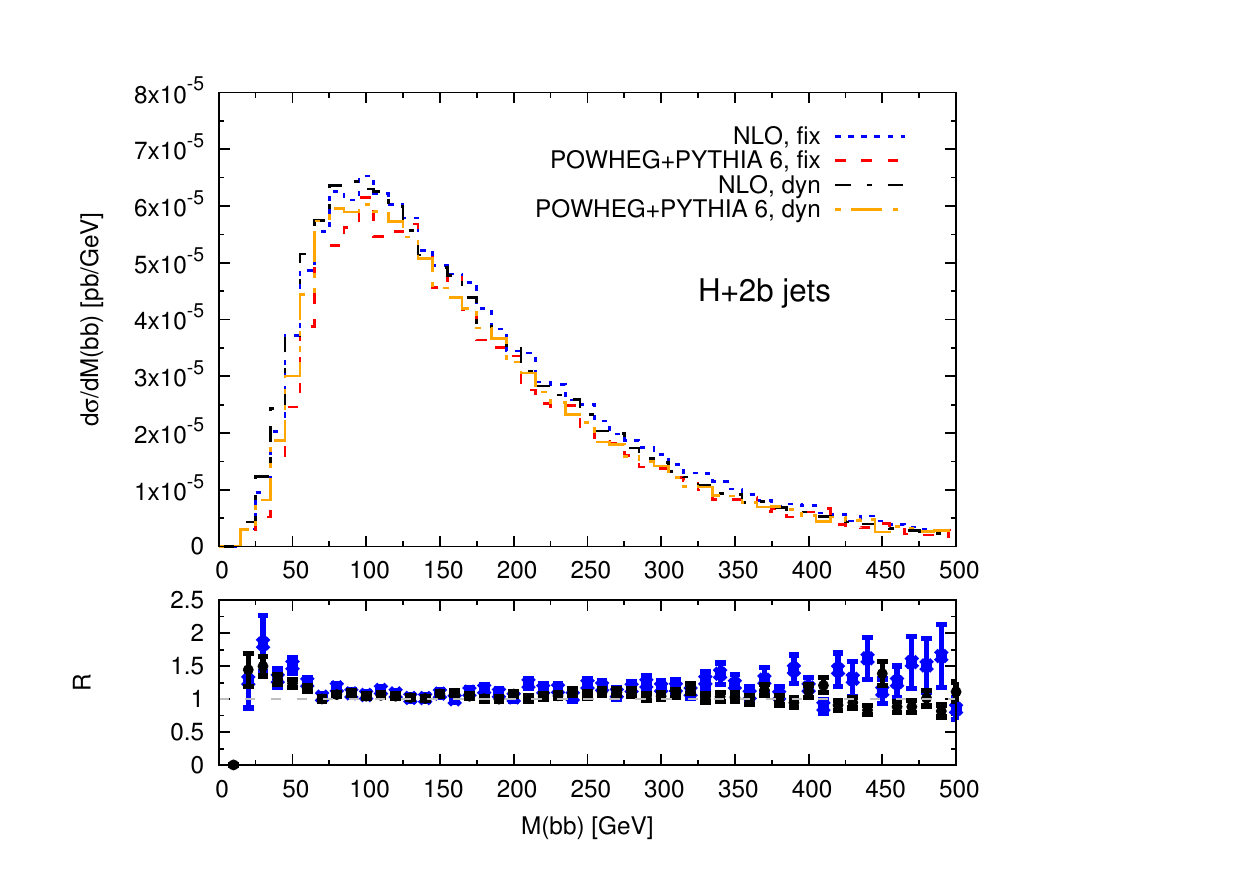} &
\includegraphics[scale=0.7,height=8truecm,width=8truecm,trim=10 0 70 20,clip]{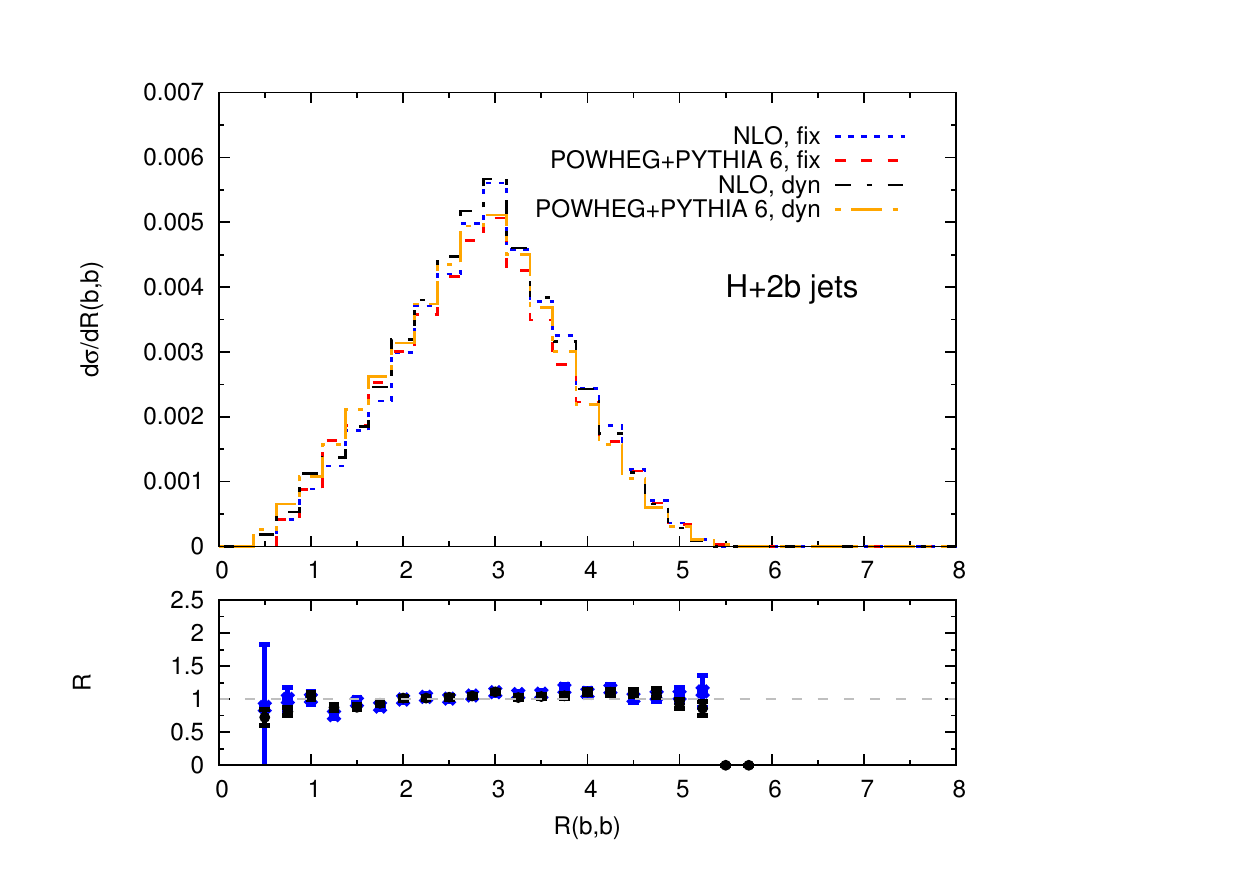}
\end{tabular}
\caption{The $M(bb)$ (left) and $R(b,b)$ (right) 
 distributions 
  in $H+2 b$-jet production at NLO-QCD with
  no parton shower (blue short-dashed line for fixed scale, black
  long-dash-dotted line for dynamical scale), and with parton shower as
  obtained through \POWHEG+\PYTHIAsix{} (red medium-dashed line for
  fixed scale, orange long-dash-dotted line for dynamical scale).  The
  lower panels show the ratios:
  $R=d\sigma(\mr{NLO})/d\sigma(\NLOPYT{})$ for a fixed (blue points)
 and a dynamical (black points) scale, respectively.  The error bars
  indicate the statistical uncertainties of the Monte-Carlo
  integration.}
\label{fig:mbbrbb_2b_nlo_pythia6}
\end{center}
\end{figure}

In Fig.~\ref{fig:mbbrbb_2b_nlo_pythia6} (left), we show the effect of
the parton shower on correlations of the two identified $b$~jets in
the $H+2 b$-jet case, in particular their invariant mass distribution
($M(bb)$) and their separation in the azimuthal-angle-pseudorapidity
plane ($R(b,b)$). The impact of the parton shower is again not
significant within the statistical errors (apart for very small
values of $M(bb)$).  

\begin{figure}[tp]
\begin{center}
\begin{tabular}{lr}
\includegraphics[scale=0.7,height=8truecm,width=8truecm,trim=10 0 70 20,clip]{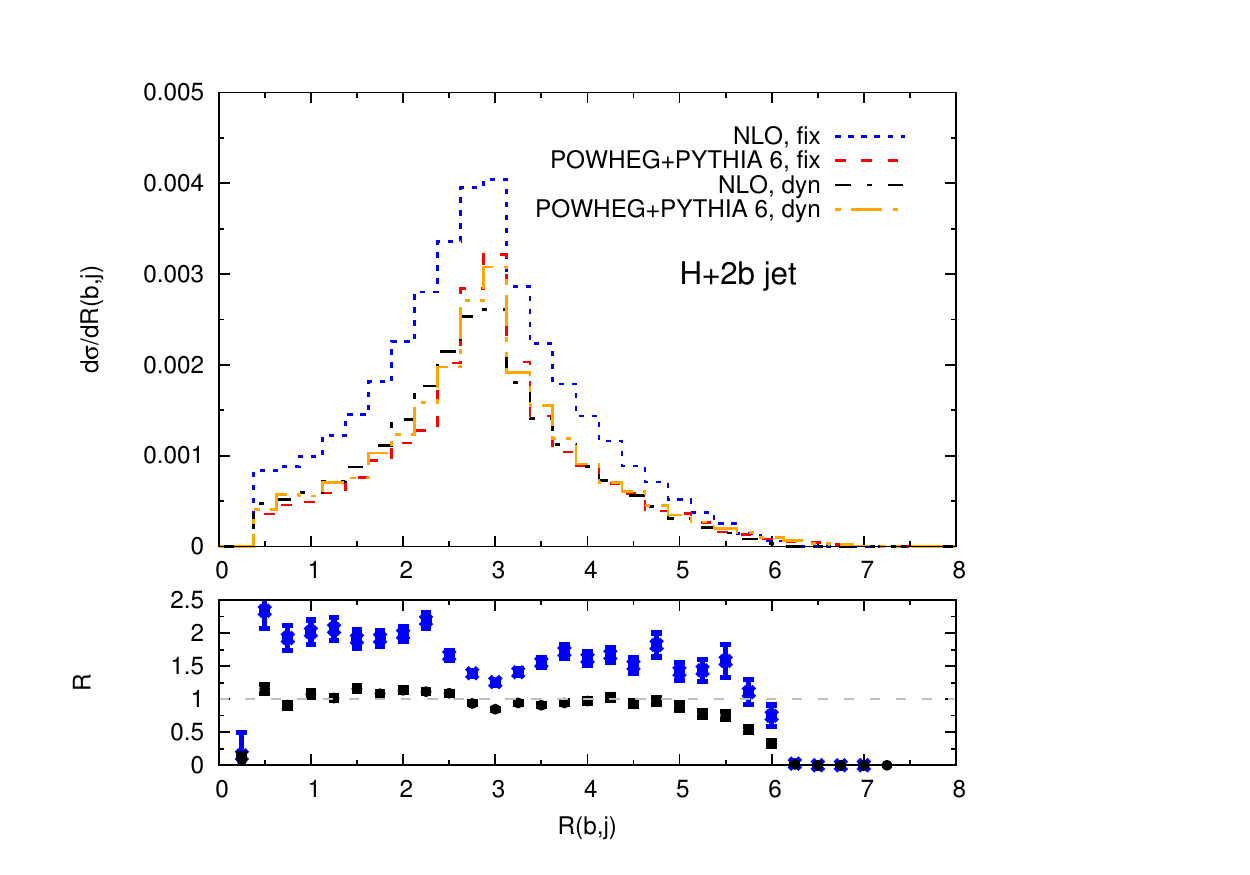}&
\includegraphics[scale=0.7,height=8truecm,width=8truecm,trim=10 0 70 20,clip]{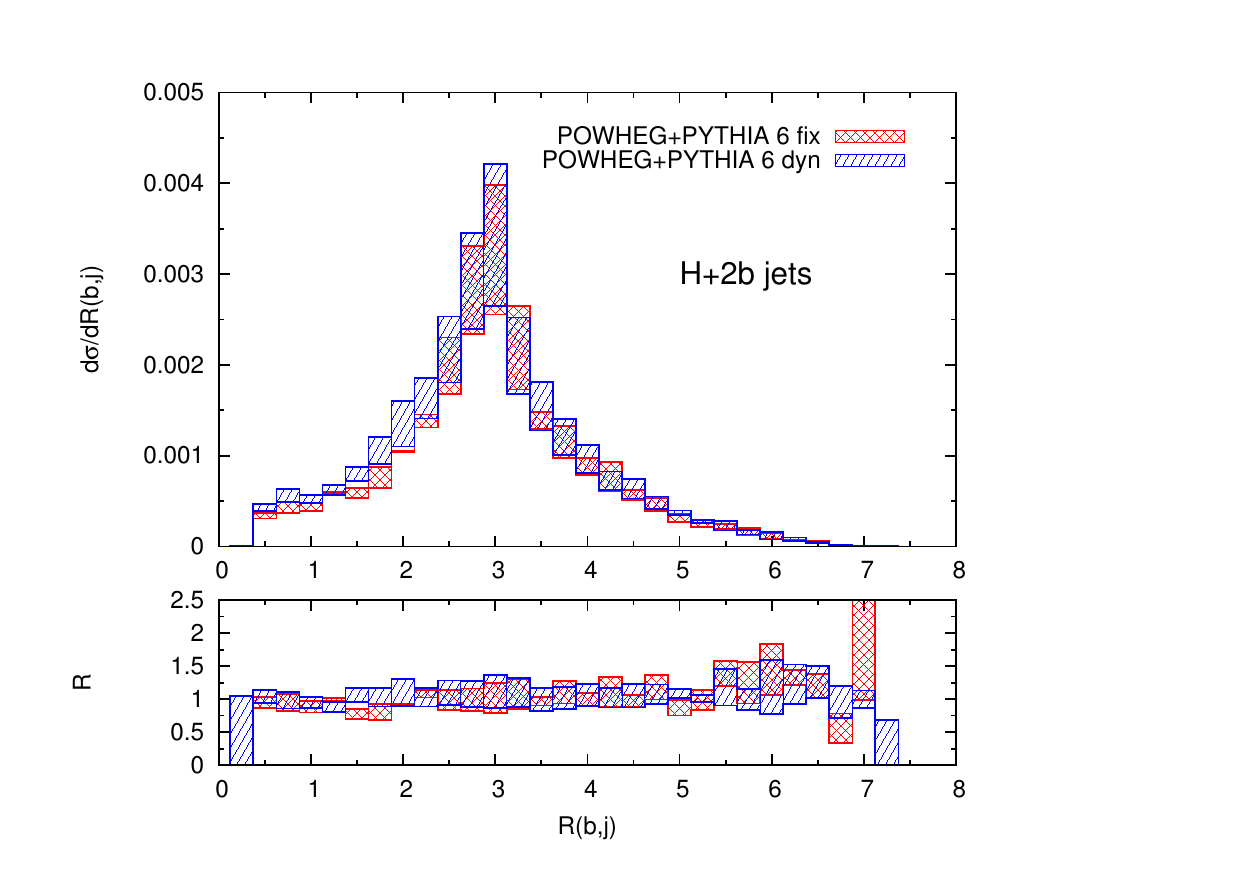}
\end{tabular}
\caption{The $R(b,j)$ distribution in $H+2 b$-jet production. The
  l.h.s. plot shows a comparison of NLO-QCD with no parton shower
  (blue short-dashed line for fixed scale, black long-dash-dotted line
  for dynamical scale), and with parton shower as obtained through
  \POWHEG+\PYTHIAsix{} (red medium-dashed line for fixed scale, orange
  long-dash-dotted line for dynamical scale).  The lower panel shows the
  ratios: $R=d\sigma(\mr{NLO})/d\sigma(\NLOPYT{})$ for a fixed (blue
  points) and a dynamical (black points) scale, respectively.  The
  r.h.s. plot shows the $R(b,j)$ NLO QCD distribution as obtained with
  \POWHEG+\PYTHIAsix{} for different values of the fixed ($fix$) and
  dynamical ($dyn$) renormalization/factorization scales, $\mu=\xi\mu_0$
  with $\xi=(0.5;2)$. The lower panel shows the respective ratios
  $R=d\sigma(\xi \mu_0)/d\sigma(\mu_0)$.  The error bars in the lower
  panels of both l.h.s. and r.h.s. plots indicate the statistical
  uncertainties of the Monte-Carlo integration.  }
\label{fig:rbj_2b_nlo_pythia6_scale_dep}
\end{center}
\end{figure}

To illustrate the behavior of distributions including a non-$b$~jet we
show in Fig.~\ref{fig:rbj_2b_nlo_pythia6_scale_dep} the $R(b,j)$
distribution in the $H+2 b$-jet case, where $R(b,j)$ is the separation
of the hardest $b$~jet and the hardest non-$b$~jet in the
azimuthal-angle-pseudorapidity plane. Here, we only consider
non-$b$~jets with a transverse momentum larger than 25~GeV in the
rapidity region of the detector $|y_j|<4.5$. In the l.h.s. plot we
show the comparison between fixed-order NLO results and results
obtained after the same calculation is interfaced with \PYTHIAsix{} in
the \POWHEGBOX{} framework, for the central value of both the fixed
and dynamical scale. Since the effect of parton shower and in particular
of scale dependence (fixed vs dynamical) seems much bigger than for
other distributions, we further investigate the scale dependence of
the distribution, which is shown in the r.h.s plot of
Fig.~\ref{fig:rbj_2b_nlo_pythia6_scale_dep}. Clearly, the $R(b,j)$
distribution is affected by a large scale uncertainty, both for a
fixed- and a dynamical-scale choice. This is typical of observables that
are indeed effectively described only at LO by a given NLO
calculation. In the case of $\bbh$ production, the hardest non-$b$~jet can only stem
from the real-emission contributions of the hard matrix element
(namely $q\bar{q},
gg\rightarrow Hb\bar{b} +g$ or $qg,\bar{q}g\rightarrow
Hb\bar{b}+q/\bar{q}$), or
from extra radiation due to the parton shower. 
Therefore, distributions of the
hardest non-$b$~jet are effectively described only with LO
accuracy and are affected by a typical LO scale uncertainty. Gaining full NLO
control on jet distributions with smaller scale uncertainties would require an NLO calculation for
$pp\to \bbh+\mr{jet}$. 
The different behavior for fixed- and
dynamical-scale choices which appears in the l.h.s. plot of
Fig.~\ref{fig:rbj_2b_nlo_pythia6_scale_dep} (where only the central
value of each scale is used) is then due to the large scale
uncertainty encountered in distributions effectively only described to
LO accuracy in the fixed-order calculation.  We find large differences
between the fixed-order predictions with $\xi = 2$ and $\xi=0.5$ when
choosing $\mur=\muf=\xi\mu_0$ with $\mu_0$ defined in
Eqs.~(\ref{eq:fix}) and (\ref{eq:dyn}). These
differences are reduced once the fixed-order calculation is combined
with the parton shower in the \POWHEG+\PYTHIAsix{} predictions.

\begin{figure}[tp]
\begin{center}
\begin{tabular}{lr}
\includegraphics[scale=0.7,height=8truecm,width=8truecm,trim=10 0 70 20,clip]{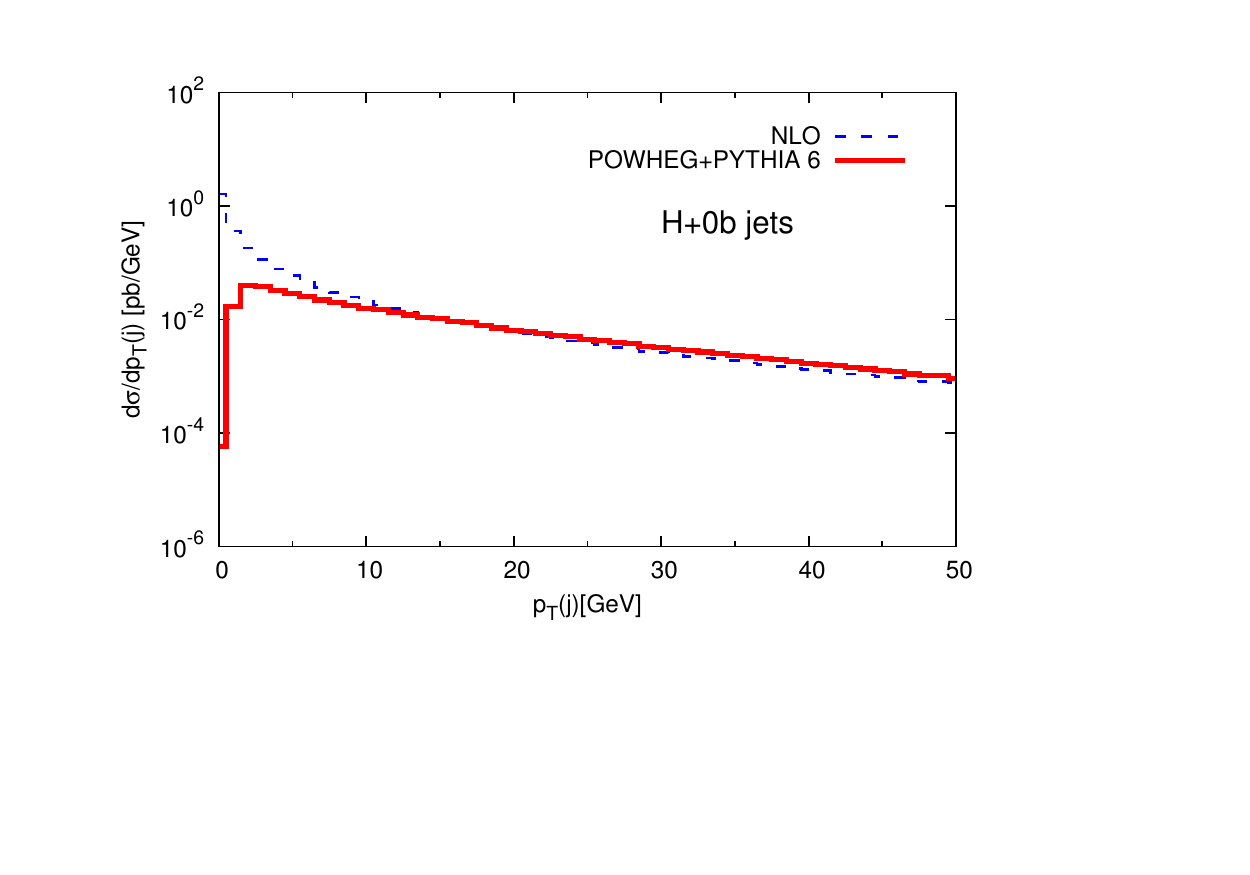} &
\includegraphics[scale=0.7,height=8truecm,width=8truecm,trim=10 0 70 20,clip]{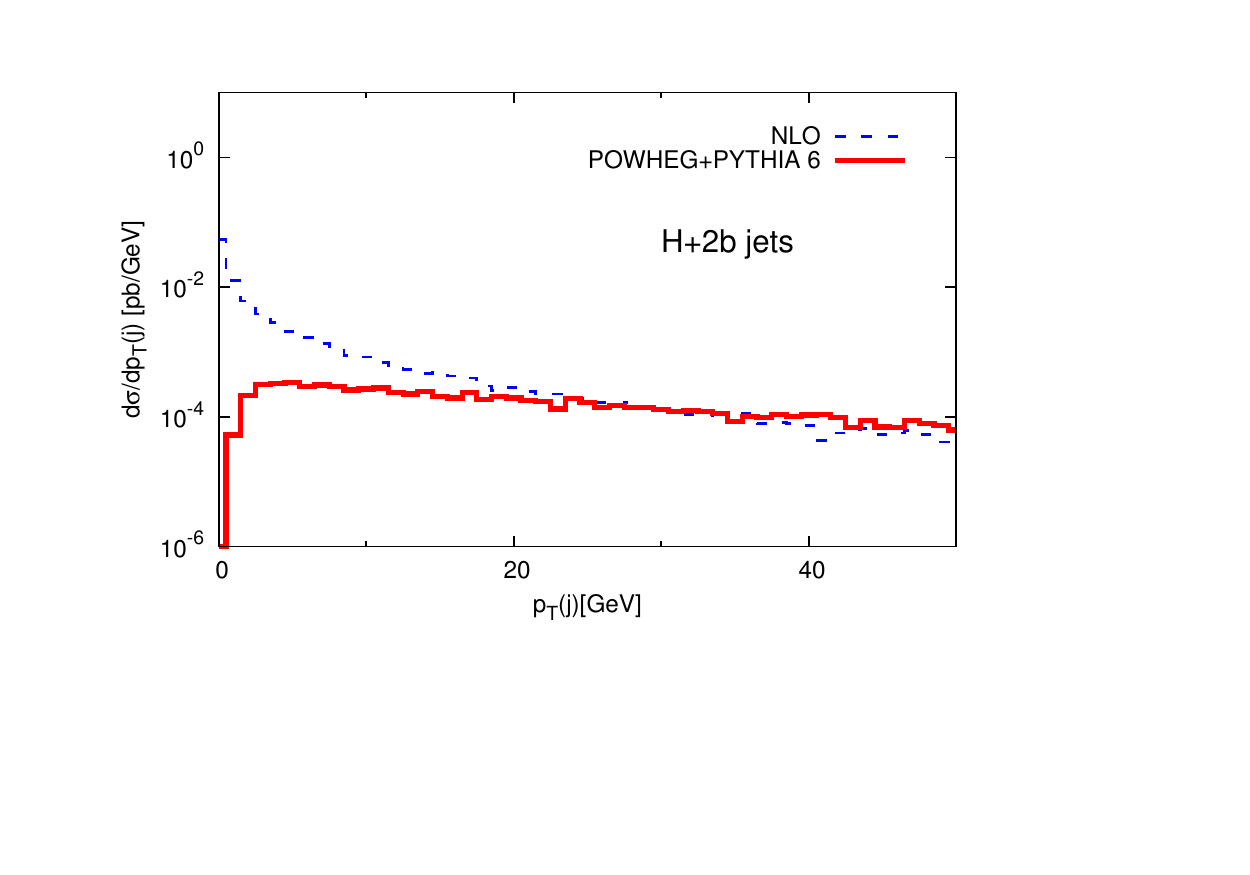} 
\end{tabular}
\caption{Transverse-momentum distribution of the hardest non-$b$ jet for the
case of inclusive (left) and $H+2 b$-jet production (right) at
  NLO-QCD with no parton shower (dashed, blue), and with parton shower
  as obtained through \POWHEG+\PYTHIAsix{} (solid, red), for a dynamical
  scale choice. }
\label{fig:jet1_nlo_pythia6}
\end{center}
\end{figure}

Indeed, to illustrate the effect of the parton shower on 
non-$b$-jet observables we show in Fig.~\ref{fig:jet1_nlo_pythia6} the transverse
momentum of the hardest non-$b$ jet, but this time with
no cut applied on the non-$b$~jet. If no $p_T$ cuts are imposed on the
non-$b$~jets, in a fixed-order calculation the transverse-momentum
distribution of a non-$b$~jet becomes very large at small values of
$p_T$, which is due to large contributions from the emission of
partons of very soft or collinear type. This behavior is tamed once
the {\POWHEG}-Sudakov factor is applied, as it is the case in the {\tt
  POWHEG+PYTHIA} result shown in Fig.~\ref{fig:jet1_nlo_pythia6} for
the case of inclusive and $H+2 b$-jet production. Since
this effect is not sensitive to the tagging of $b$~jets, similar
shapes are encountered in both analysis scenarios (and also in the 
$H+1 b$-jet scenario which is not explicitly shown here).


\begin{figure}[tp]
\begin{center}
\begin{tabular}{lr}
\includegraphics[scale=0.7,height=8truecm,width=8truecm,trim=10 0 70 20,clip]{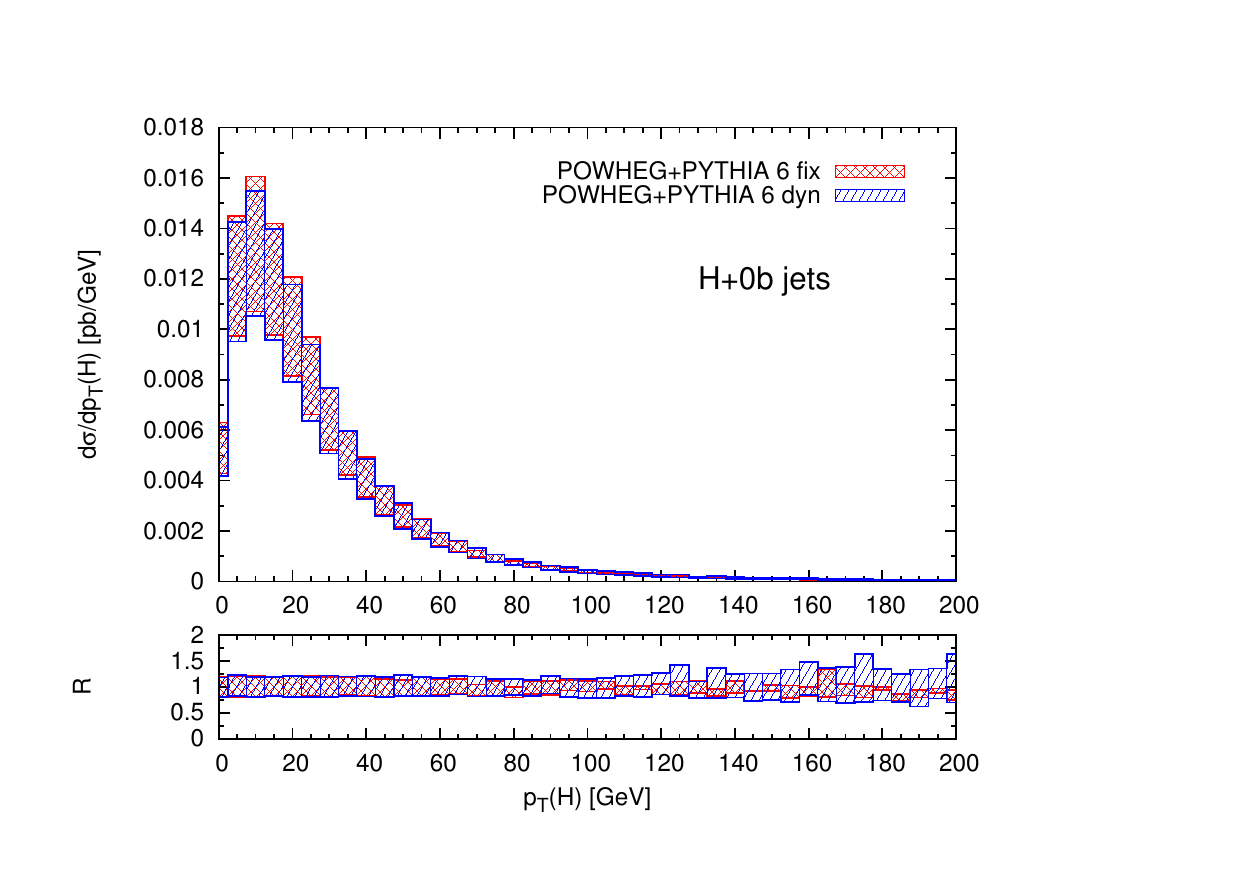}&
\includegraphics[scale=0.7,height=8truecm,width=8truecm,trim=10 0 70 20,clip]{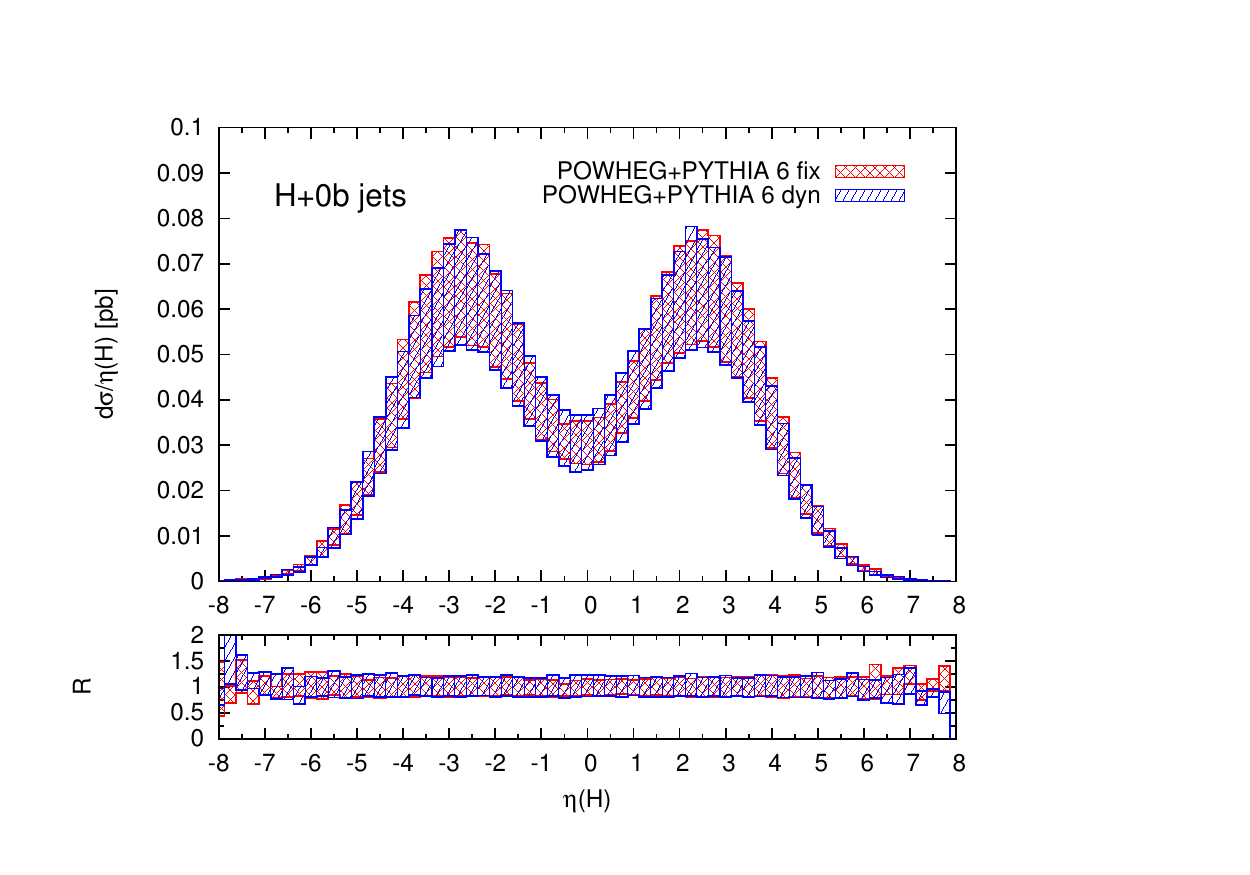}
\end{tabular}
\caption{The $p_T$ (left) and $\eta$ (right) distributions of the
  Higgs boson in the inclusive-production case as obtained with
  \POWHEG+\PYTHIAsix{} for different values of the fixed ($fix$) and
  dynamical ($dyn$) renormalization/factorization scales,
  $\mu=\xi\mu_0$ with $\xi=(0.5;2)$. The lower panels show the
  respective ratios $R=d\sigma(\xi\mu_0)/d\sigma(\mu_0)$.}
\label{fig:pteta_h_scale_dep_0b}
\end{center}
\end{figure}

\begin{figure}[tp]
\begin{center}
\begin{tabular}{lr}
\includegraphics[scale=0.7,height=8truecm,width=8truecm,trim=10 0 70 20,clip]{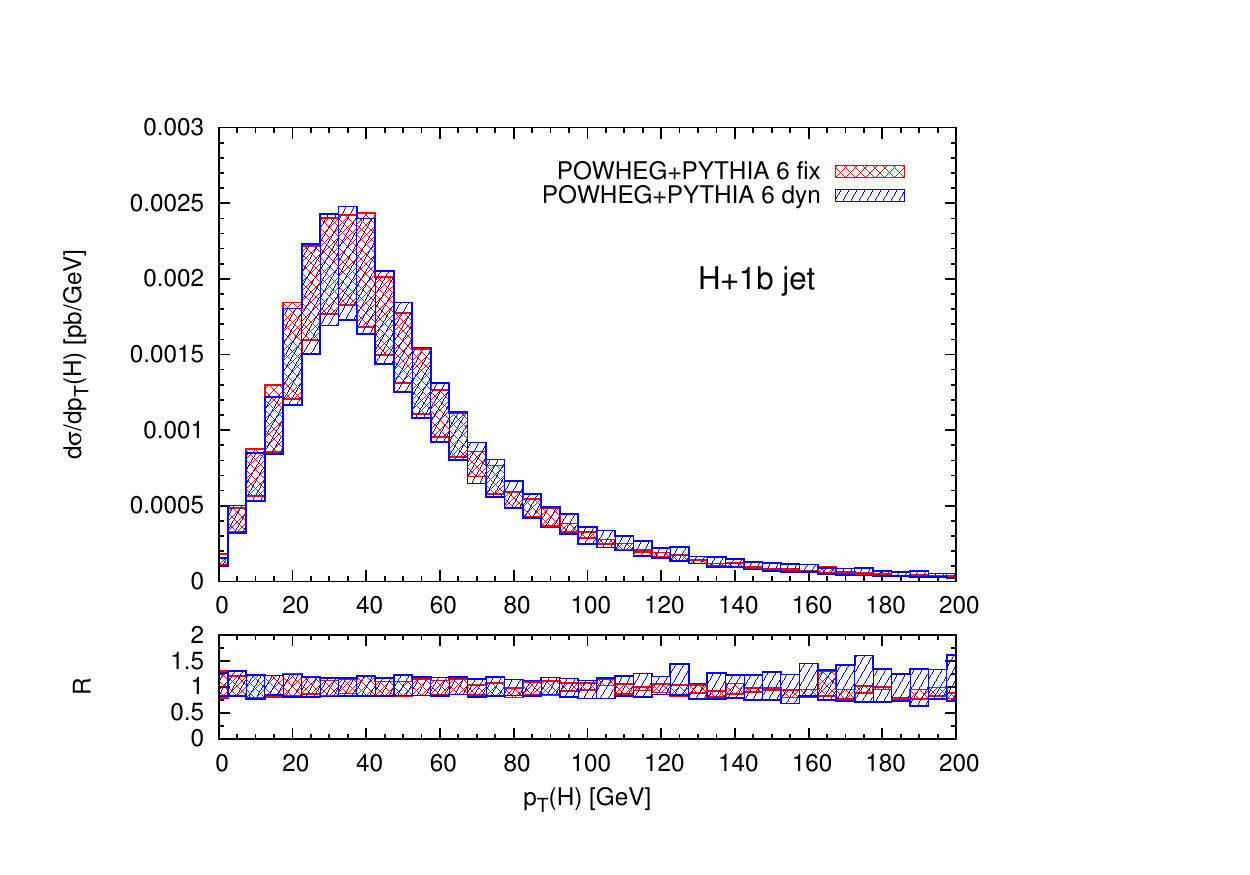}&
\includegraphics[scale=0.7,height=8truecm,width=8truecm,trim=10 0 70 20,clip]{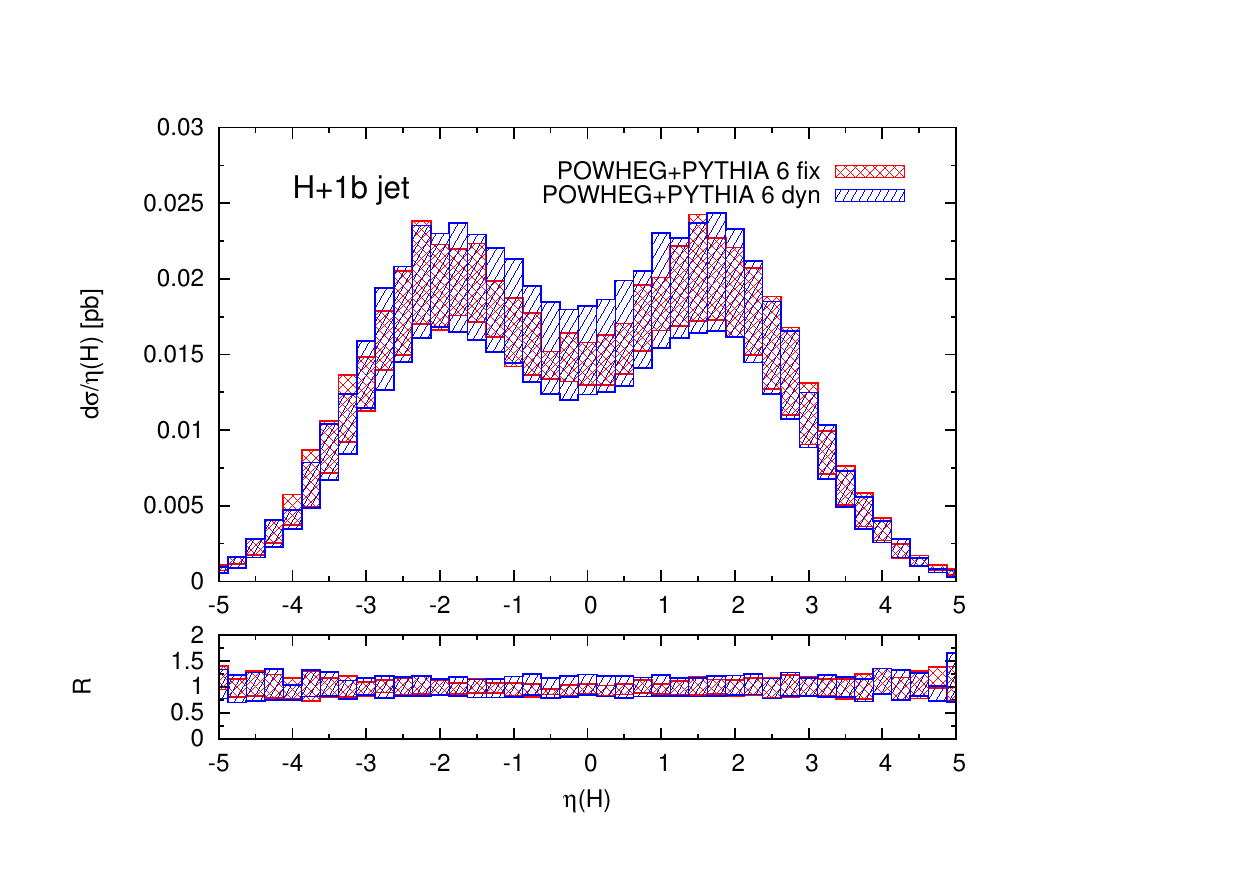}
\end{tabular}
\caption{The $p_T$ (left) and $\eta$ (right) distributions of the
  Higgs boson in $H+1b$-jet production as obtained with
  \POWHEG+\PYTHIAsix{} for different values of the fixed ($fix$) and
  dynamical ($dyn$) renormalization/factorization scales,
  $\mu=\xi\mu_0$ with $\xi=(0.5;2)$. The lower panels show the
  respective ratios $R=d\sigma(\xi\mu_0)/d\sigma(\mu_0)$.}
\label{fig:pteta_h_scale_dep_1b}
\end{center}
\end{figure}

\begin{figure}[tp]
\begin{center}
\begin{tabular}{lr}
\includegraphics[scale=0.7,height=8truecm,width=8truecm,trim=10 0 70 20,clip]{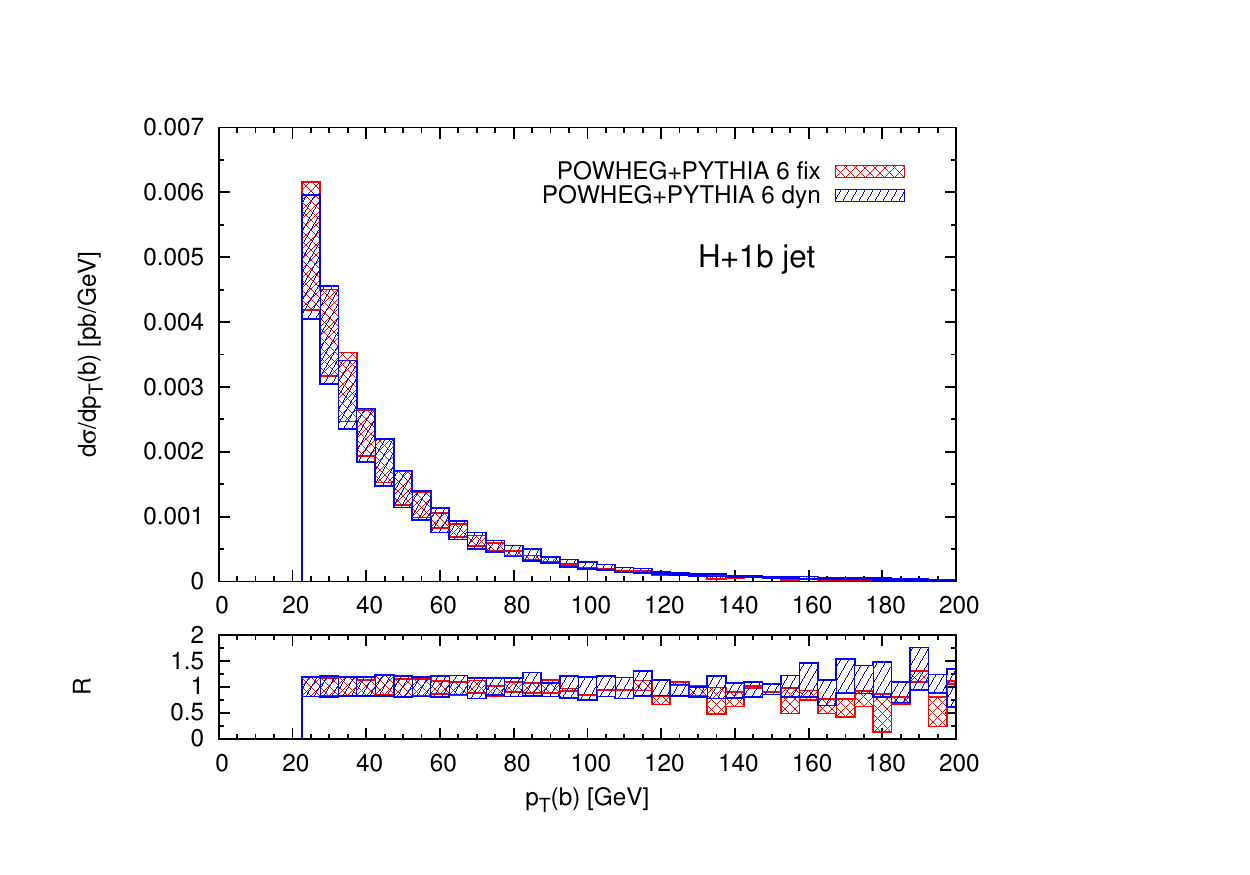}&
\includegraphics[scale=0.7,height=8truecm,width=8truecm,trim=10 0 70 20,clip]{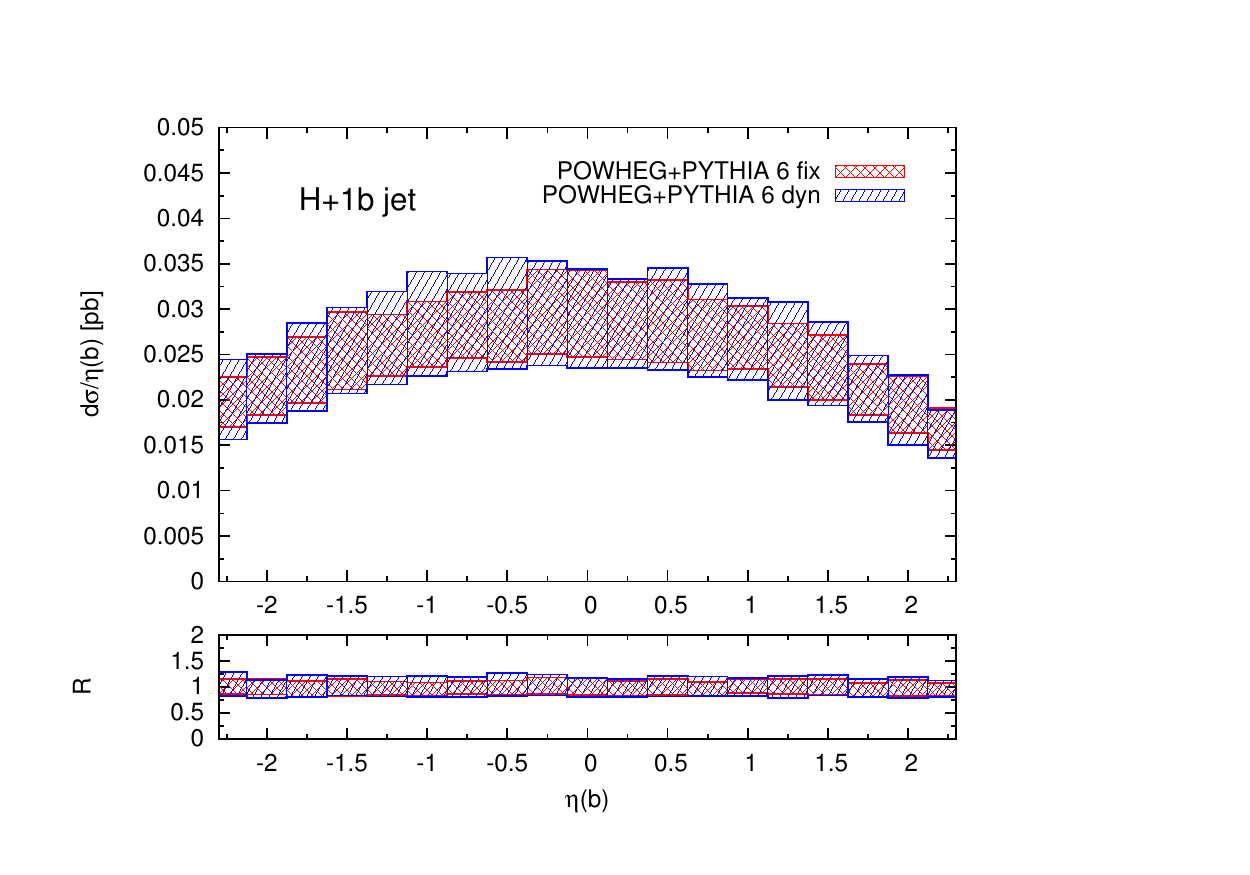}
\end{tabular}
\caption{The $p_T$ (left) and $\eta$ (right) distributions of the
  hardest identified $b$~jet in $H+1b$-jet production as obtained with
  \POWHEG+\PYTHIAsix{} for different values of the fixed ($fix$) and
  dynamical ($dyn$) renormalization/factorization scales,
  $\mu=\xi\mu_0$ with $\xi=(0.5;2)$. The lower panels show the
  respective ratios $R=d\sigma(\xi\mu_0)/d\sigma(\mu_0)$.}
\label{fig:pteta_b_scale_dep_1b}
\end{center}
\end{figure}

\begin{figure}[tp]
\begin{center}
\begin{tabular}{lr}
\includegraphics[scale=0.7,height=8truecm,width=8truecm,trim=10 0 70 20,clip]{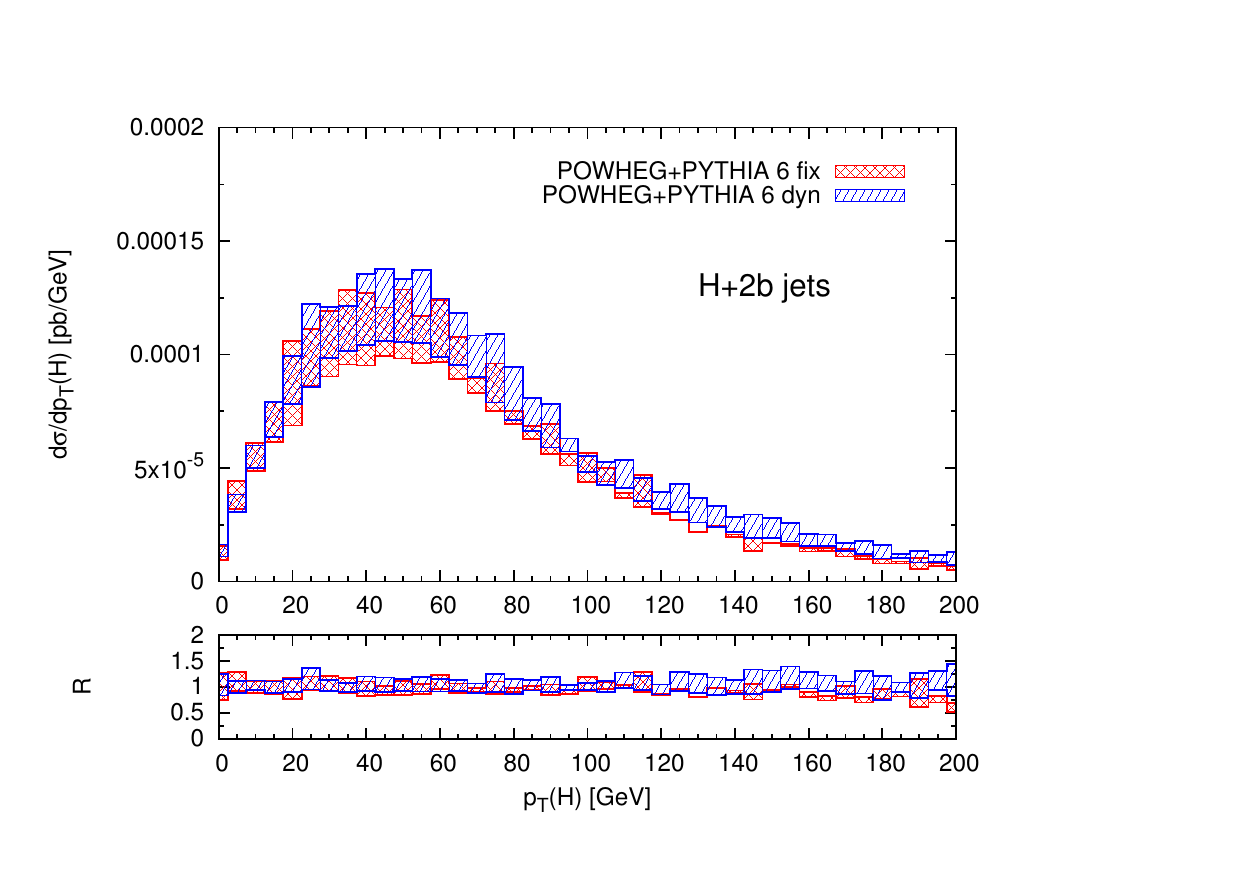}&
\includegraphics[scale=0.7,height=8truecm,width=8truecm,trim=10 0 70 20,clip]{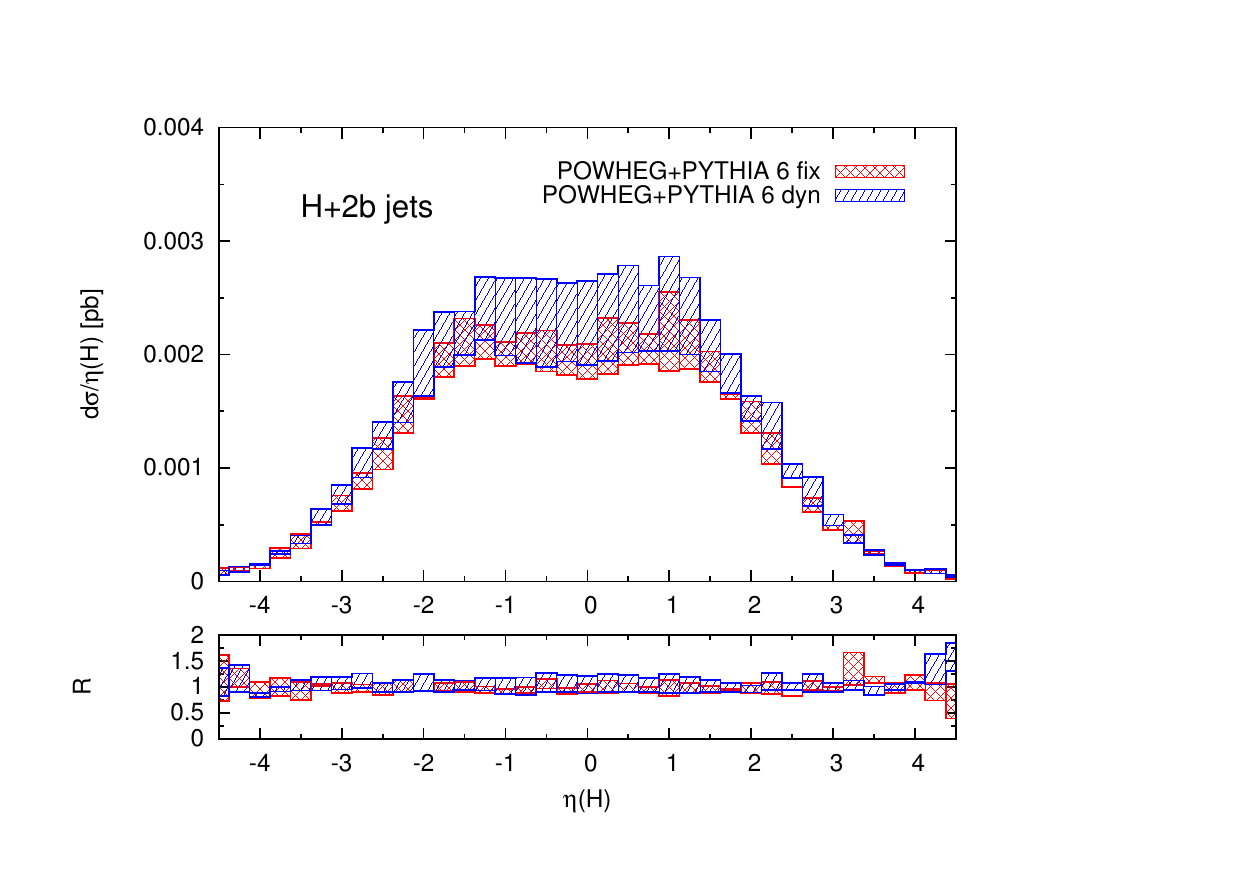}
\end{tabular}
\caption{The $p_T$ (left) and $\eta$ (right) distributions of the
  Higgs boson in $H+2b$-jet production as obtained with
  \POWHEG+\PYTHIAsix{} for different values of the fixed ($fix$) and
  dynamical ($dyn$) renormalization/factorization scales,
  $\mu=\xi\mu_0$ with $\xi=(0.5;2)$. The lower panels show the
  respective ratios $R=d\sigma(\xi\mu_0)/d\sigma(\mu_0)$.}
\label{fig:pteta_h_scale_dep_2b}
\end{center}
\end{figure}

\begin{figure}[tp]
\begin{center}
\begin{tabular}{lr}
\includegraphics[scale=0.7,height=8truecm,width=8truecm,trim=10 0 70 20,clip]{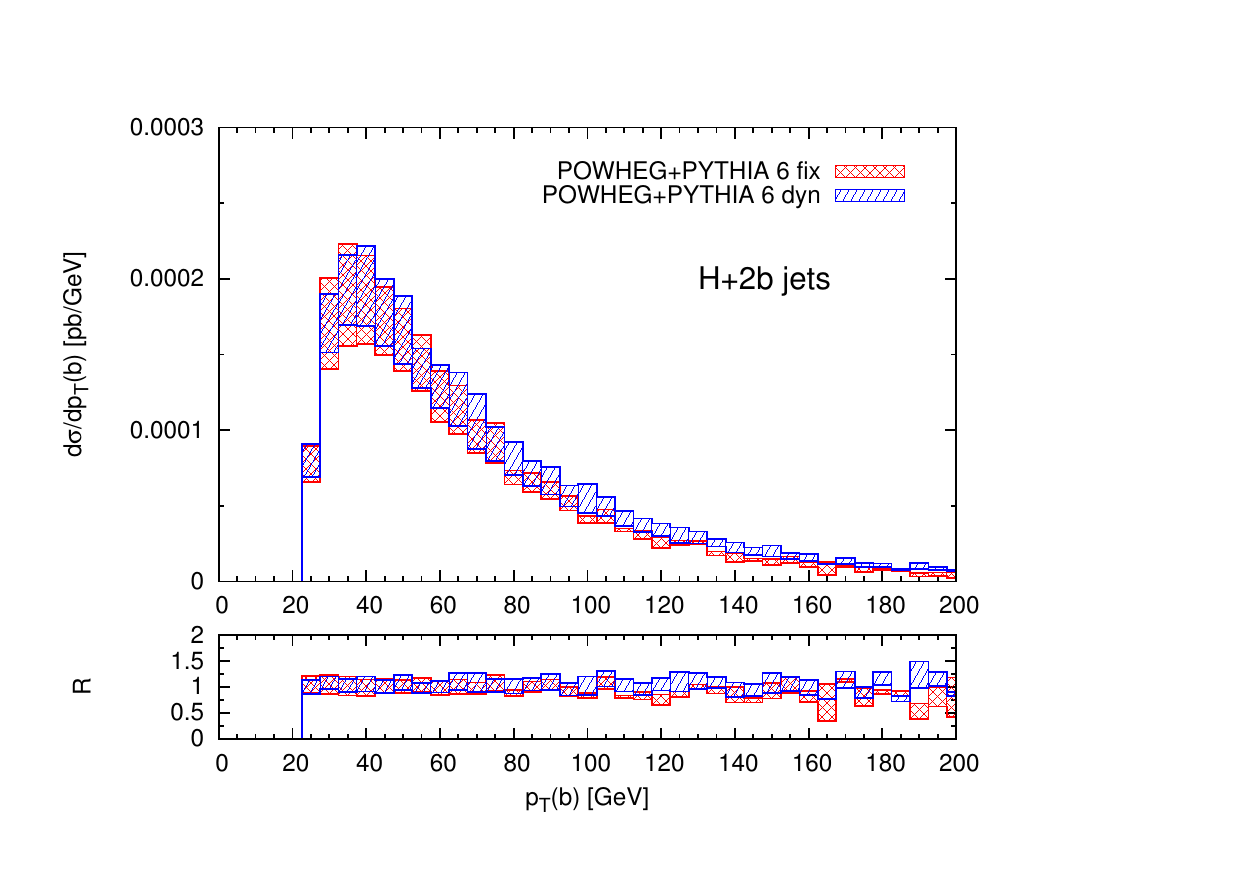}&
\includegraphics[scale=0.7,height=8truecm,width=8truecm,trim=10 0 70 20,clip]{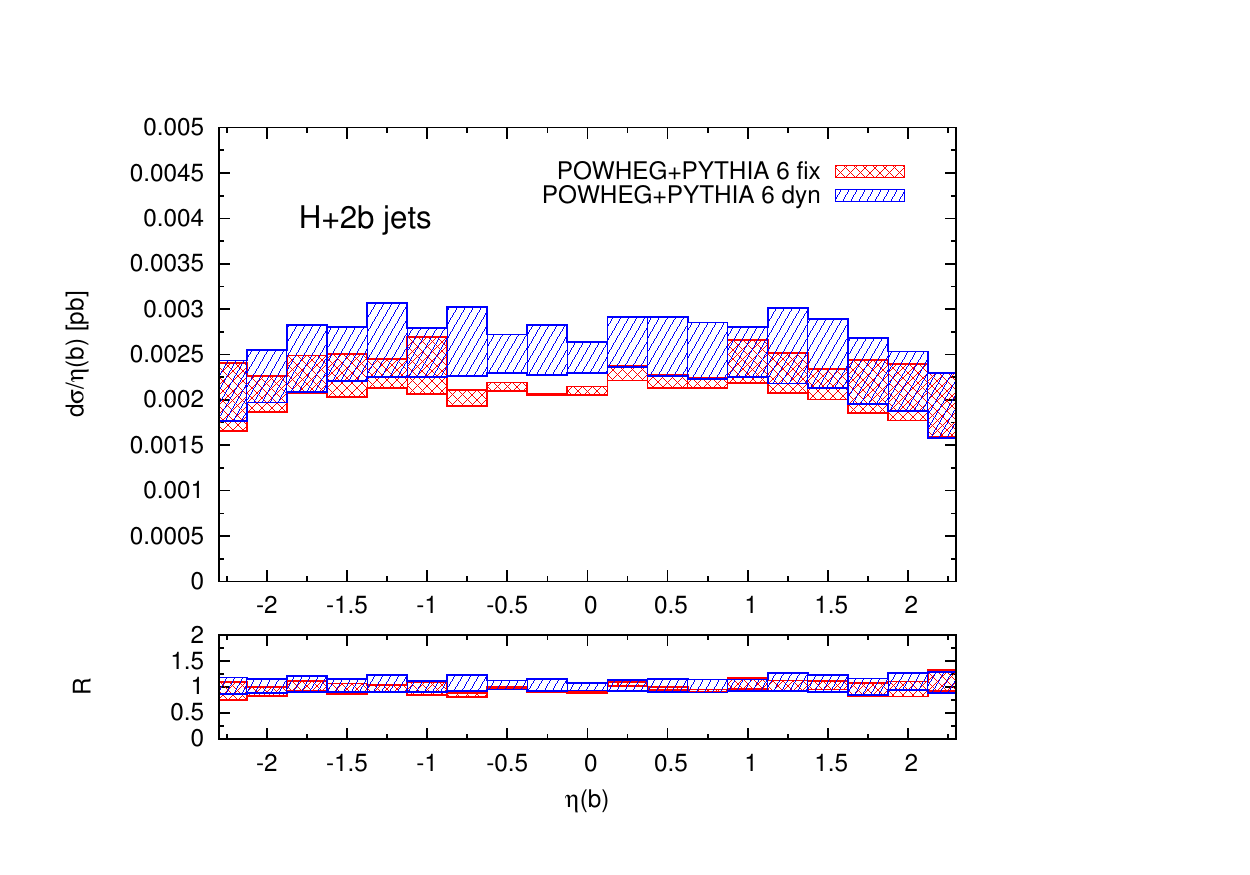}
\end{tabular}
\caption{The $p_T$ (left) and $\eta$ (right) distributions of the
  hardest identified $b$~jet in $H+2b$-jet production as obtained with
  \POWHEG+\PYTHIAsix{} for different values of the fixed ($fix$) and
  dynamical ($dyn$) renormalization/factorization scales,
  $\mu=\xi\mu_0$ with $\xi=(0.5;2)$. The lower panels show the
  respective ratios $R=d\sigma(\xi \mu_0)/d\sigma(\mu_0)$.}
\label{fig:pteta_b_scale_dep_2b}
\end{center}
\end{figure}

%
%
In order to assess the theoretical uncertainties associated with the
choice of renormalization and factorization scale for NLO
distributions, we have computed the $p_T$ and $\eta$ distributions of
the Higgs boson for all three signatures, i.e. $H+0,1,2 b$~jets, and
the $p_T$ and $\eta$ distributions of the hardest identified $b$~jet
for both $H+1b$-jet and $H+2b$-jet production, for different choices
of scales as discussed earlier. The corresponding results, obtained
using our \POWHEG+\PYTHIAsix{} implementation, are shown in
Fig.~\ref{fig:pteta_h_scale_dep_0b} (no $b$ tagging),
Figs.~\ref{fig:pteta_h_scale_dep_1b}-\ref{fig:pteta_b_scale_dep_1b}
($H+1 b$~jet), and
Figs.~\ref{fig:pteta_h_scale_dep_2b}-\ref{fig:pteta_b_scale_dep_2b}
($H+2 b$~jets).  The scale dependence of the results is considerable,
amounting to about $\pm 25\%$ in most regions of phase space. Using a
fixed scale rather than a dynamical scale helps in slightly reducing
the scale uncertainty of the NLO+\PYTHIA{} results in all observables,
especially at larger values of $p_T(b)$, $p_T(H)$ and in the central
pseudorapidity region, although the effect is moderate. We note that
also the total cross sections in the three scenarios considered here
exhibit a large scale uncertainty, for instance we find $\sigma=0.477
\, {\rm pb} \pm 18\%$ for the total inclusive cross section obtained
with our NLO+\PYTHIAsix{} implementation in the setup of
Fig.~\ref{fig:pteta_h_scale_dep_0b} for a fixed-scale choice.


\section{Conclusions}
\label{sec:conclusions}
In this article we have presented the implementation of the NLO QCD
calculation for $\bbh$ production at a hadron collider (from
Ref.~\cite{Dawson:2003kb}) in the \POWHEGBOX{} package. We emphasize
how having $H+b$-jet production available in the \POWHEGBOX{} provides
a crucial element of consistency for experimental studies that rely on
the same framework for a broad variety of signal and background
processes.  The code is made publicly available so that it can be used for
further studies of $\bbh$ production at the LHC in the SM and in
extensions of the SM with modified Yukawa couplings of the third
generation quarks. Here, we considered the SM and presented numerical
results at fixed perturbative order and at NLO QCD matched with
\PYTHIA{} for selected representative setups. In particular, we
discussed theoretical uncertainties due to factorization/renormalization scale 
choices and variations and parton-shower effects for two
analysis scenarios with a Higgs boson in association with one or two
$b$~jets, respectively. A sample of results for the case of inclusive production
have also been presented. We found that parton-shower effects do not
give rise to large distortions of observables related to the Higgs
boson or identified $b$~jets in $H+1 b$-jet and $H+2 b$-jet production
processes at the LHC. As expected, more pronounced effects occur in
distributions involving non-$b$ jets, as shown, for example, by 
the transverse-momentum distribution of the hardest non-$b$~jet or by the
distribution of the separation ($R(b,j)$) between the hardest $b$ jet
and the hardest non-$b$ jet.  We studied the
impact of different scale choices on various distributions and found
that the associated theoretical uncertainties can be considerable.

\section*{Acknowledgements}
We are grateful to Carlo Oleari for his assistance in making this code
publicly available on the \POWHEGBOX{} website.  The work of B.~J.~is
supported in part by the Institutional Strategy of the University of
T\"ubingen (DFG, ZUK~63) and in part by the German Federal Ministry
for Education and Research (BMBF) under contract number 05H2015.  The
work of L.~R.~is supported in part by the U.S. Department of Energy
under grant DE-FG02-13ER41942.  The work of D.~W. is supported in part
by the U.S. National Science Foundation under award no.~PHY-1118138.

\bibliography{bbh}

\end{document}